\documentclass[aps]{emulateapj}
\usepackage{amsmath,mathtools,float,subfigure}
\usepackage{color,ulem,epstopdf}
\usepackage{graphicx}

\def\refnew#1{(\ref{#1})}
\newcommand{\K}{\rm \, K}
\newcommand{\g}{\rm \, g}
\newcommand{\cm}{\rm \, cm}

\shorttitle{Mass and Scalings for Super-Earths}
\shortauthors{Wu, Yanqin}

\begin{document}

\title{Mass and Mass Scalings of Super-Earths}
% ubiquitous, pervasive, natural, prevalent, 

\author{Yanqin Wu\altaffilmark{1}}
\affil{Department of Astronomy and Astrophysics, University of Toronto, Toronto, ON M5S 3H4, Canada}
\email{wu@astro.utoronto.ca}

\begin{abstract} 
The majority of the transiting planets discovered by the {\it Kepler} mission (called super-Earths here, includes the so-called 'sub-Neptunes') orbit close to their stars. As such, photoevaporation of their hydrogen envelopes etch sharp features in an otherwise bland space spanned by planet radius and orbital period.  This, in turn, can be exploited to reveal the mass of these planets, in addition to techniques such as radial velocity and transit-timing-variation.  Here, using updated radii for {\it Kepler} planet hosts from Gaia DR2, I show that the photoevaporation features shift systematically to larger radius for planets around more massive stars (ranging from M-dwarfs to F-dwarfs), corresponding to a nearly linear scaling between planet mass and its host mass.  By modelling planet evolution under photo-evaporation, one further deduces that the masses of super-Earths peak narrowly around $8 M_\oplus (M_*/M_\odot)$. { When such a stellar mass dependence is scaled out, {\it Kepler} planets appear to be  a homogeneous population surprisingly uniform in mass, in core composition (likely terrestrial) and in initial mass fraction of their H/He envelope (a couple percent).}
% Moreover, the composition of their cores is likely terrestrial, and they were initially coated %with H/He envelopes a couple percent in mass. 
%Interestingly, 
The masses of these planets do not appear to depend on the metallicity values of their host stars, while they may depend on the orbital separation weakly.
% possibly as $r^{1/2}$. 
Taken together, the simplest interpretation of our results is that super-Earths are at the so-called 'thermal mass', where the planet's Hill radius is equal to the vertical scale height of the gas disk.  \end{abstract}


\section{Introduction}

%{\w have yet to show that mu for actie disk is mass-indep.}

%{\w some comments from my email box:

%  You say that there is no statistical test for 2D. But I don't see the difference between 1D and 2D (or n dimensions). If you know the error of individual data points, you can always compute a chisqr for example. Am I missing something here?  

%Teske: -How exactly do you calculate the Gaia DR2 stellar radii? 
%-Do you recalculate the Kepler planet radii based on the improved stellar radii? 
%-Why do you restrict yourself to "naked core" planets in Figure 8 and the associated
%argument? Shouldn't the entire small planet density/mass distribution peak where you
%expect it to, since your model spans all small planets? 

%Li Zeng: I just read your new paper on ArXiv (https://arxiv.org/abs/1806.04693 [arxiv.org] ). 
%Your Figure 11 is in agreement with our earlier work:  especially the gap dependence as
%stellar mass/radius to the 1/4-th power
% “Exoplanet Radius Gap Dependence on Host Star Type”. Li Zeng, Stein B. Jacobsen,
%Dimitar D. Sasselov. Research Notes of the AAS (RNAAS), 2017.
%(http://iopscience.iop.org/article/10.3847/25155172/aa9ed9/meta [iopscience.iop.org] )

%daniel huber: Gaia parameters are optimized all-sky while ours use the large
%amount of follow-up done for the Kepler field, so we think our values are superior to
%those in the Gaia catalog for Kepler targets.
%Our catalog used in the current arxiv paper version is available here:

%https://drive.google.com/open?id=1SMPs8vXvNGVdxZNaaCdKtXtmxH-q97LH %[drive.google.com]
%}

%{\w Kepler planets are a uniform cohort.}

Close-in, low-mass planets (which we call 'super-Earths', which also include the so-called 'sub-Neptunes') appear commonly. Earlier statistics \citep{Petigura,Fressin} suggest that some $60\%$ of all stars possess them, while updated modelling that takes into account planet multiplicity reduces this frequency to $30\%$ { for solar-type hosts} \citep{Zhu18}. This is 2-3 times more common than that of giant planets.  We are keenly interested in knowing the masses of these planets and how these masses depend on parameters like stellar mass, stellar metallicity, and orbital distances.

% a paragraph on mass determinations
Nature provides us with two well-known ways of measuring planet masses, the radial velocity technique \citep[e.g.,][]{mayor-small,marcy} and the tool of transit-timing-variation \citep[e.g.,][]{HolmanMurray,Agol,Lithwick12}. Both have produced a large body of results. Unfortunately, for the majority of the super-Earths known, neither is available.

% a paragraph on photo-evaporation, how it works and how it is confirmed
At first glance, these planets appear to be a diverse population, with a range of transiting sizes. However, it was pointed out early on that, their size distribution is bi-modal, with smaller planets occupying shorter periods, and larger ones at longer periods, and that this is likely the signature of photoevaporation \citep{WuLithwick}.  Over time, refined stellar parameters firmly established the bi-modal distribution \citep{Fulton,vaneylen}, and model developments \citep{Lopez,OwenWu13, OwenWu17, JinMordasini} confirmed the conjecture of photoevaporation.  Of particular relevance to this work, \citet{OwenWu17} developed a simple semi-analytical model to efficiently follow the evaporation and thermal evolution of low-mass planets. This model showed that the timescale for mass-loss is the longest where planet sizes (core plus H/He envelope) are doubled from their core sizes. For a range of parameters, this naturally gives rise to a final planet distribution that is bimodal in radius, peaking at the naked core size and twice its value. This explains the steep fall-off of planets beyond $\sim 2.6 R_\oplus$, and a gap around $1.8 R_\oplus$. The stunning evaporation valley, evident in the recent CKS data \citep{Fulton}, gives solid proof of this theory.
%Put it another way, planets with sizes that differ by a factor of 2 actually have the same mass.

% a paragraph on the intention of this work
Confirmation of the photoevaporation model now provides a new tool for us to measure planet mass and composition. This was attempted preliminarily in \citet{OwenWu17}, where the core composition was inferred to be Earth-like, and the core mass is shown to have a variance of $3 M_\oplus$. We revisit this problem and provide a more in-depth, updated analysis.

The current work is made possible by advances on two fronts. First, a reliable list of planets are now carefully vetted from thousands of candidates \citep{Borucki11, Batalha13,Burke14,Mullally15,Rowe15}.
% on Kepler confirmed planets, lots of vetting
Second, large surveys like the California Kepler Survey \citep{Petigura-cks,Johnson-cks}, the LAMOST Kepler survey \citep{Dong-lamost}, and the Gaia mission have provided reliable stellar radii, which lead to better determined planet radii.
% on Gaia new data, how is stellar radii obtained
In particular, the Gaia Collaboration recently released radii and effective temperatures for most of the {\it Kepler} planet hosts. This latter release forms much of the basis for this work.

\section{Gaia DR2 data}

Newly released Gaia data (DR2) provide stellar parameters for most of the {\it Kepler} planet host stars, based on Gaia photometry and parallaxes \citep{gaia_brown,Lindegren18,Andrae18}.  We retrieve a list of 2296 confirmed {\it Kepler} planets from the NASA Exoplanet archive \citep{Akeson13} in May 2018.  We query the Gaia database for objects within $1"$ of the positions of the host stars, requiring that the matching pairs be within $1$ magnitude of each other (Gaia broad-band magnitude is similar to Kepler broad-band magnitude in wavelength range). We find Gaia radii for 1551 stars, which host altogether 2157 confirmed {\it Kepler} planets.
% statistics from sm.s/merge_gaia
This sample is larger than the CKS catalogue \citep{Petigura-cks,Johnson-cks}, which contains $1298$ confirmed planets.
 % 2025-727 
More importantly, unlike the CKS sample that focus on FGK dwarfs, the Gaia sample contains stars  from M-dwarfs to F-dwarfs.

How accurate are the Gaia stellar radii?  One empirical test is the sharpness of the so-called 'photoevaporation valley' in the radius-period plane. With the original KIC data, this is buried in the noise; data from spectroscopic surveys like CKS allow us to see it clearly in the first time, for a subset of the {\it Kepler} planets \citep{Fulton}. With Gaia radii, the valley is prominent in the overall sample (left panel of Fig. \ref{fig:Rad-period-2}), as well as in individual groups. The 1-D size distribution one obtains using the Gaia radii has features as sharp as those obtained using the CKS data. 

Recently, \citet{FultonPetigura} computed a new set of stellar radii for the CKS sample, by combining high resolution spectroscopy and Gaia astrometry. These agree closely with the asteroseismically determined values whenever available, with an RMS dispersion of $3.4\%$ in the radius ratio. Comparing these with the Gaia DR2 radii, they found that the latter are in average $0.4\%$ larger, with a $6.1\%$ RMS dispersion in the radius ratio. This dispersion is tolerable for our purpose.\footnote{Another set of radius determination, based off Gaia parallaxes, is provided recently by \citet{Berger}. Data is not yet public as the paper is still under review. It is unclear how different they are from the Gaia values.}
% Bergert+18: To calculate stellar radii we used the stellar classi- fication code isoclassify (Huber et al. 2017) in %its “direct method”, using as input the Gaia DR2 parallax (Lindegren et al. 2018), 2MASS K-band magnitude, %and Teff , log g, [Fe/H] values from the DR25 KSPC (Mathur et al. 2017). ; Updated for CKS stars; and Teff < %4000K by Gaidos+16; Gaia parameters are optimized for > 160 million stars across the %sky. In contrast, the %Kepler field is one of the most well-studied samples of stars due to its relevance to %exoplanet science, and %the KSPC in- cludes information from the vast amount of photometric, spectroscopic %and asteroseismic %analyses that have been performed over the past ten years. Therefore, we expect the stellar radii derived in %this work are preferable over those reported by the Gaia Collaboration.
So in this work, we adopt the Gaia radii and consider it to have an uncertainty of $7\%$.

Here, we also use Gaia stellar radii as a proxy for stellar mass.  Within a factor of two of solar-mass, main-sequence radius can be well approximated as $R_* = R_\odot (M_*/M_\odot)$ \citep[see review by][]{Torres}.
%ZAMS:  0.2 - 1.4 M_sun, M ~ R^0.9; 1.4-30 M_sun, M ~ R^0.6a
There is one subtlety, however. Based on Gaia radius and $T_{\rm eff}$, some of the planet hosts have clearly evolved and are onto the sub-giant stage. So we enforce a cut to reject all stars with
\begin{equation}
R_* \geq 2.0 R_\odot \times \left({{T_{\rm eff}}\over{6000\K}}\right)^{1.8}\, .
\label{eq:mscut}
\end{equation}
This removes $316$ planets, and we are left with a total of $1841$. We call this the Gaia-Kepler sample, or GKS for short.

\begin{figure} 
\includegraphics[width=0.45\textwidth,trim=20 150 60 100,clip=]{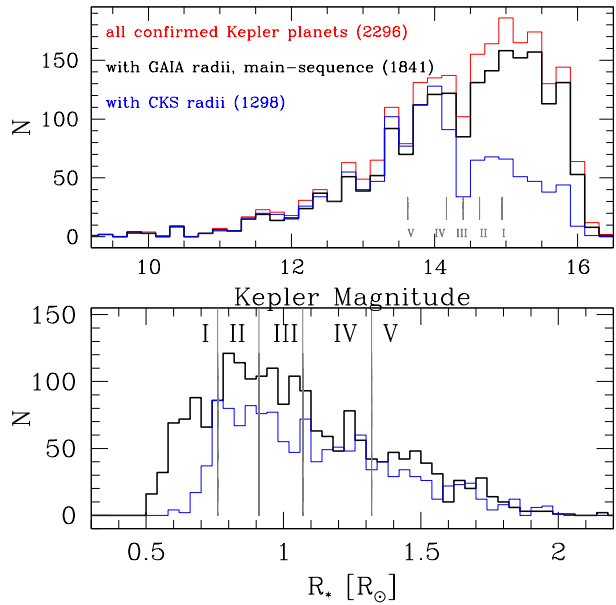}
\caption{Host star properties for the Gaia-Kepler planet sample. The upper figure shows the distribution of host-star magitudes for all {\it Kepler} confirmed planets, those with Gaia radius determinations (the GKS), and those with CKS determinations. The GKS is nearly complete at all magnitudes, while the CKS sample focus on brighter stars. The lower figure compares the ranges of stellar radius coverage between the GKS and the CKS samples -- the former spans a larger range, and included stars down to $R = 0.5 R_\odot$ (early M-dwarfs).
% M0 at 0.63 R_sun, M_0=0.47 M_sun
We divide the GKS into five equal groups at stellar radii $0.76 R_\odot, 0.91 R_\odot, 1.07 R_\odot, 1.32 R_\odot$ (vertical lines in lower panel). These groups span from M-dwarfs, to K-, G-, and F-dwarfs. Different groups have different mean magnitudes, and are marked by short vertical bars in the top panel. These affect planet detectability and should be accounted for when modelling the planet population.
% average magnitude of Gaia sample is 14.35. Groups V-I are 13.62, 14.17, 14.4, 14.63, 14.94
% dimmer stars 
}
\label{fig:gaia-property} 
\end{figure}

We now divide GKS planets into five equal groups by stellar radius, as is illustrated in Fig. \ref{fig:gaia-property}. The median stellar radii for these groups span from $0.67 R_\odot$ to $1.5 R_\odot$, i.e., a dynamic range of $\sim 2.3$ in stellar radius (and mass).\footnote{See Fig. \ref{fig:Rad-period-mass} for an even larger dynamic range.}

Fig. \ref{fig:Rad-period-mass} provides an overview of these planets, plotted as planet radius versus orbital period, separately for the five groups of host stars. As stars become more massive, planets in general become larger, and in particular, features in the radius-period plane (caused by photo-evaporation and discussed below) systematically move upward. In addition, there is a slight shift of the features toward longer periods.

Could this be caused by the selection effect where small planets are harder to detect around larger stars? In the following, we discuss observational selections and reject this hypothesis. Instead, we argue, the difference is physical.

% sm.s/gaia_property

\subsection{Detection Efficiency}
\label{subsec:selection}

Transiting planets are detected subject to their transit probability, which is simply $R_*/a$. But even where they transit, the detection efficiency still lies below unity, as the transit signal is not guaranteed to be picked up by the pipe-line.

%the average Kepler magnitude is 14.4 mag, which is close to the average magnitudes of entire Kepler planet %sample (14.34).

For an average {\it Kepler} star (Kepler magnitude $14.34$, $R_* \approx R_\odot, M_* \approx M_\odot$), the
% actually mean radius is 1.05 R_\odot, sm.s/gaia_property
detection completness is established by \citet{Burke15,Christiansen15}.
% --- argued to be flawed by Dong & Xie... 
The curves delineating various completeness levels lie roughly at planet radius \citep[as summarized in Fig. 1 of][]{Zhu18}
\begin{equation}
  R_{14.34} =
\begin{dcases}
& 1.07 R_\oplus \left(P\over{1 {\rm days}}\right)^{0.2}\,, \hskip0.2in  90\% {\,\, \rm complete}\, \nonumber \\
 & 0.67 R_\oplus \left(P\over{1 {\rm days}}\right)^{0.2}\,, \hskip0.2in  50\% {\,\, \rm complete}\, \nonumber \\
  & 0.42 R_\oplus \left(P\over{1 {\rm days}}\right)^{0.2}\,, \hskip0.2in 10\% {\,\, \rm complete}\,.
\end{dcases}
\label{eq:completeness}
\end{equation}
The above period scaling is similar to, within uncertainty, the analytical scaling of $P^{1/6}$ derived in  \citet{GaidosMann}.

For a general star of a radius $R_*$, and Kepler magnitude $K_p$, we follow the analytical arguments in  \citet{GaidosMann} to write
\begin{equation}
R_{R_*,K_p}  = R_{14.34} \left({{R_*}\over{R_\odot}}\right)^{3/4} \left({{M_*}\over{M_\odot}}\right)^{1/12} \times 10^{{{K_p-14.34}\over{5}}}\, .
\label{eq:gaidosmann} 
\end{equation}
%Gaidos & Mann provides a scaling R_min ~ p^{1/6} d^{1/2}, where d is distance
%set diff=(100.0**(1./5.)) set diff=sqrt(diff)
% this is difference in distance between stars that have a difference of one magnitudes in brightness, 
%set y1=y/diff**0.5 set y2=y/diff**1.0 set y3=y/diff**1.5 set y4=y/diff**2.0

\begin{figure*} 
\includegraphics[width=\textwidth,trim=0 0 0 0,clip=]{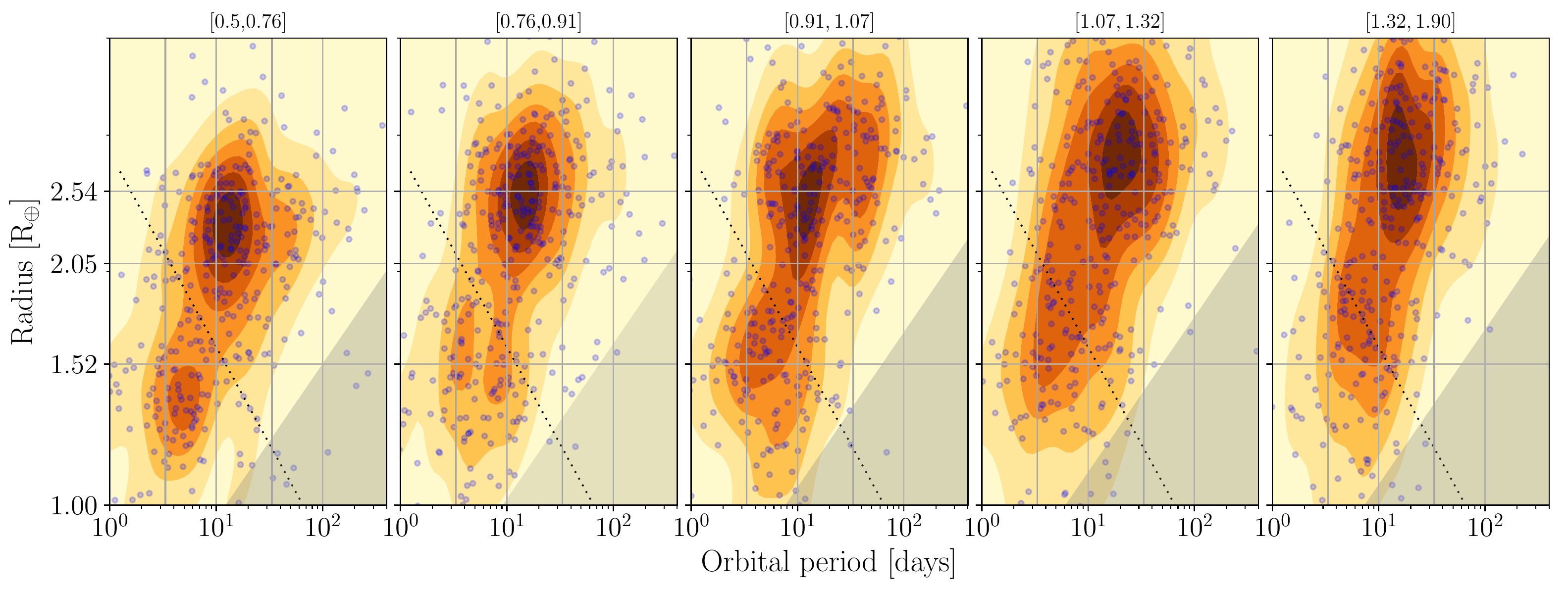}
\caption{Planet radius vs orbital period, for all five groups in GKS. The dotted lines describe the rough 
location of the photoevaporation valley and are mostly for guidance. Features in the plane progressively move upward for more massive stars. The shaded areas indicate regions below which detection completeness falls below $50\%$ (assuming transiting) for different groups of stars. These look similar for all groups -- larger stars tend to appear brighter (Fig. \ref{fig:gaia-property}) and this compensates for the rise in their 
%$14.94, 14.63, 14.4, 14.17,13.62$, 
radii. The main observed features all lie well above where completeness matters. }
\label{fig:Rad-period-mass} 
\end{figure*}

%As the {\it Kepler} field probes somewhat above the disk plane, most stars are at distances 1kpc or %smaller. And as a result, intrinsically brighter stars appear also brighter.
% the Sun M=4.83, at d=1kpc, m=14.83, 
% F0 star, M=3.0, m=13; G0 star, M=4.7,m=14.7
% K0 star, M=6.0, m=16.0; M0 star, m=18.7
While small planets are harder to see around larger stars, for our five groups of stars, the
{ fact that the} average brightness increases with stellar radius largely compensates { for this}. { Planets of smaller (relative) radii can be detected around stars that are brighter due to the higher signal-to-noise ratio.} As a result, we find similar completeness curves for all of them (Fig. \ref{fig:Rad-period-mass}). These curves fall well below the observed features in the radius-period plane.  So the prominent radius inflation we observe in that figure is unlikely to be due to selection bias.

\subsection{Planet mass Scales with Stellar Mass}

\begin{figure*} 
\includegraphics[width=0.45\textwidth,trim=20 130 50 120,clip=]{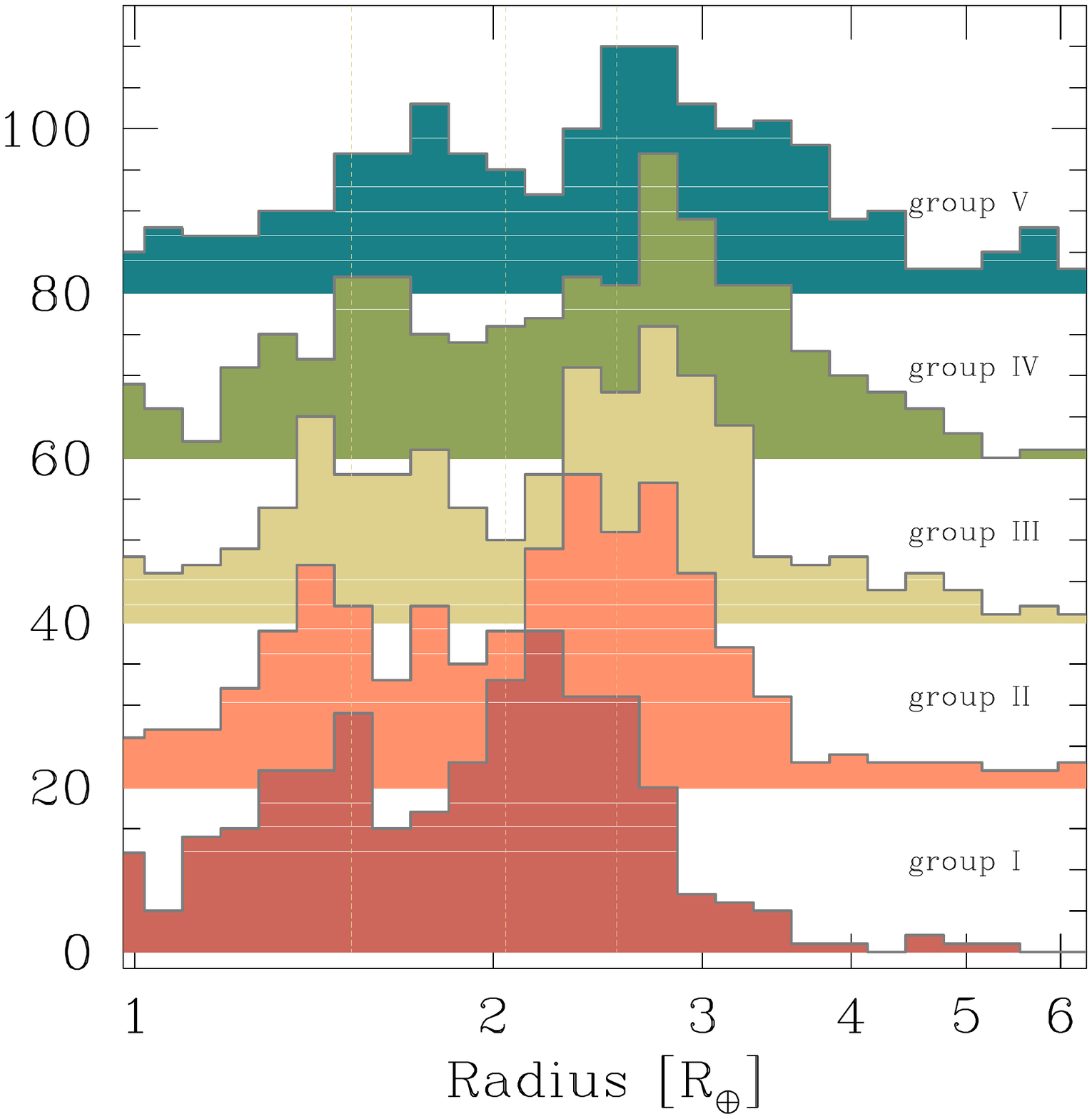}
% if want to use the pretty one, use gaia_radius_shifted0.pdf
\includegraphics[width=0.45\textwidth,trim=20 130 50 120,clip=]{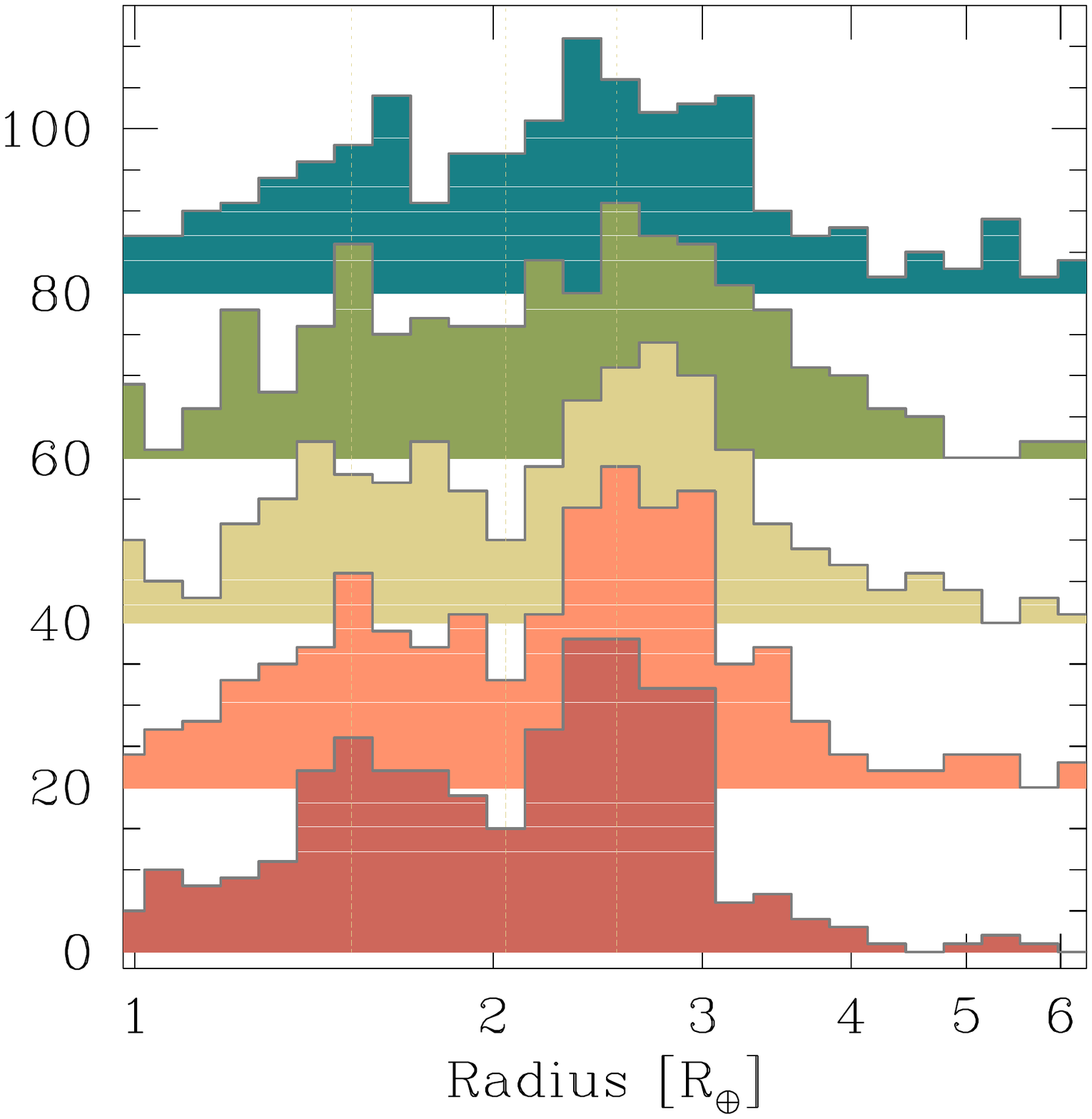} 
\caption{Left panel: 1-D radius distribution for the five groups of planets, vertically displaced for clarity.  The resolution of the histogram is chosen to be $7\%$ of the radius, the presumed error in Gaia radius determination.  The olive vertical lines are drawn to show the locations of ``features'' for planets around group III stars { (the middle group)}: first peak ($R = 1.52 R_\oplus$), dip ($2.05R_\oplus$) and second peak ($2.54 R_\oplus$). As the stellar mass increases (bottom to top), the same features systematically shift to the right.  Right panel: the same distributions but for the scaled radius, $R_p' = R_p (M_*/M_\odot)^{-1/4}$. Such a mass scaling not only lines up the features across different groups, it also sharpens the features within the same group.  } \label{fig:gaia-property-2} \end{figure*}

Similar to what is observed in 2-D space (Fig. \ref{fig:Rad-period-mass}), the 1-D radius distributions for our groups I-V stars also exhibit the same progressive shift ( Fig. \ref{fig:gaia-property-2}). As stars become more massive, the features (peaks and dip) move systematically to the right.  What could cause these upward shifting patterns? Here, we use photoevaporation as a tool to argue that this is caused (solely) by the rise of planet mass with stellar mass.

Consider a homogeneous population of planets, similar in core mass, core composition, envelope mass, thermal history, and host star UV/X-ray irradiation, and different only in their distances from the host star. Photo-evaporation will rob the close-in ones of their envelopes, leaving them as naked cores, while the very far-away planets will largely retain their envelopes. Planets at intermediate distances, on the other hand, will preferentially retain envelopes of a few percent in mass (and appearing at twice the core sizes), as the timescale to evaporate is the longest at these sizes\citep{OwenWu17}. This produces the so-called photo-evaporation valley in the radius-period plane, and in the 1-D size distribution, a pattern that has two peaks, one at the core size, one at roughly twice that.  The detection of these patterns in the {\it Kepler} sample, in both the 2-D radius-period plane, and in the 1-D radius histogram, not only confirms the theory of photoevaporation, but can be put into a more profitable use.

If the planet population around different stars are the same, the peaks and dips should occur at the same radii for all stars. On the other hand, if the characteristic planet mass ($M_0$) scales with stellar mass ($M_*$) as
\begin{equation}
M_0 \propto M_*^\beta \, ,
\label{eq:M0}
\end{equation}
we expect to see a synchronous progression of the radius features with stellar mass, as is
observed in Fig. \ref{fig:Rad-period-mass} (2-D) and Fig. \ref{fig:gaia-property-2}.

To ascertain the value of $\beta$, we note that core sizes scale with core masses as $R_p \propto M_p^{1/4}$ \citep{fortney2007a}, for a given chemical composition. So if we scale the planet size appropriately as $R_p' = R_p (M_*/M_\odot)^{-\beta/4}$, there will be a value of $\beta$ at which the peaks and dips in the (scaled) 1-D size distributions coincide across all groups of stars. At this point, the peak-to-dip ratio in the combined histogram will have the largest value, or, the image is the sharpest.

\begin{figure} \includegraphics[width=0.45\textwidth,trim=20 220 150 120,clip=]{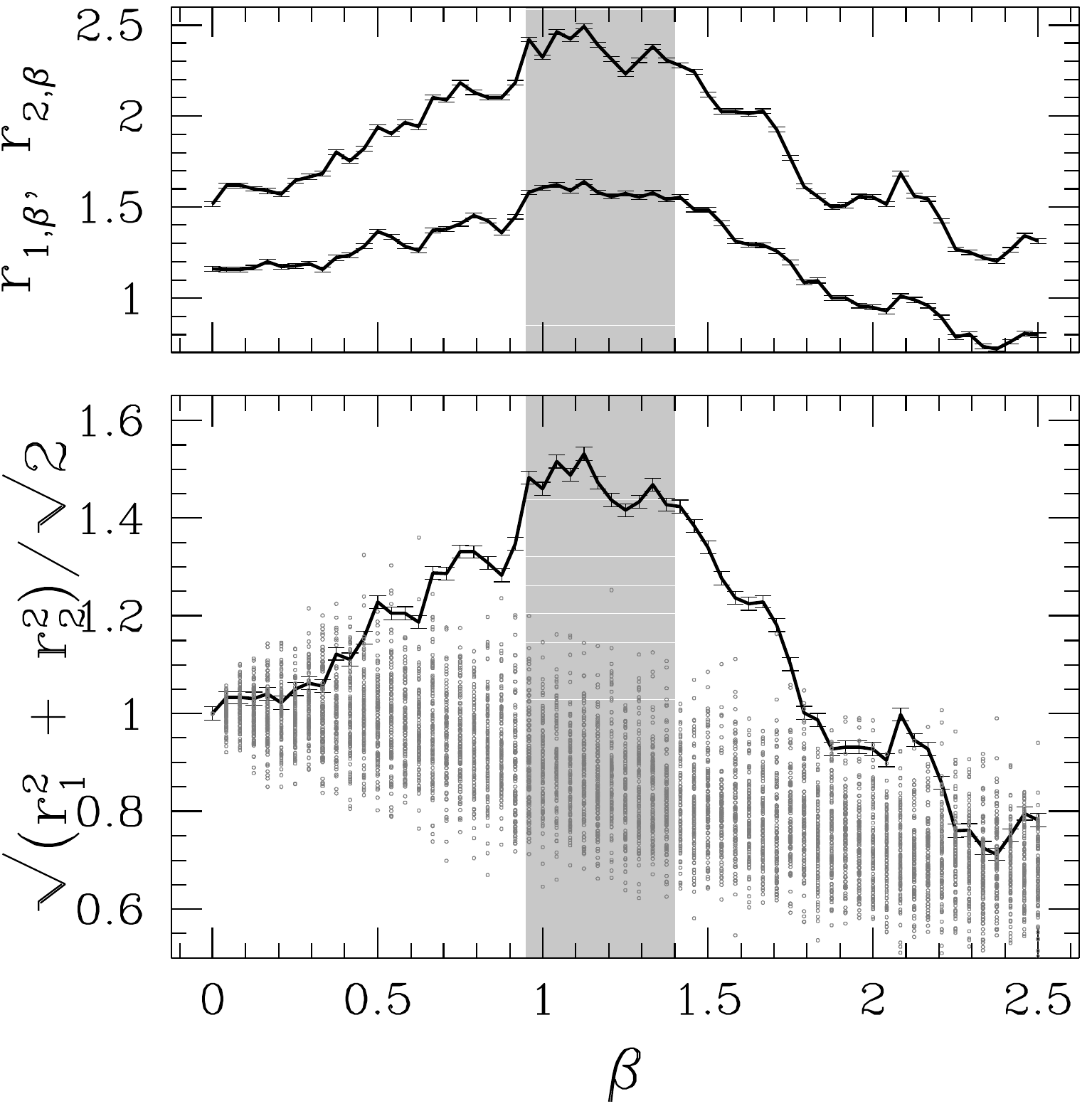} \caption{Changes in the peak-to-dip ratios in the size histogram as the planet mass (and radius) is scaled by various values of $\beta$ (eq. [\ref{eq:M0}]). The top panel shows the two peak-to-dip ratios, $r_{1,\beta}${ (lower curve)} and $r_{2,\beta}$ { (top curve)}, together with their poisson errors.  A choice of $\beta \approx 1$ maximizes both ratios.  The lower panel shows a combination of these two ratios, $\sqrt{r_1^2 + r_2^2}/\sqrt{2}$, where $r_1 = r_{1,\beta}/r_{1,0}$, $r_2 = r_{2,\beta}/r_{2,0}$, which is also maximized at $\beta\approx 1$. We mark with the 
shaded grey zone roughly the range of acceptable $\beta$, judged by twice the poisson errors.
To argue that this is not a chance event, we overplot (in grey  points) results of $100$ runs where the planet-star associations are scrambled randomly. None of the runs produce as large a peak-to-dip ratio (i.e., as sharp features) as that at $\beta\approx 1$ for the real run. }
\label{fig:gaia-shift-radius} 
\end{figure}

Using a logarithmic radius bin of $7\%$, we produce combined histograms of $R_p'$ for all groups. We then use the heights at the following three bins: $R_1 = 1.52 R_\oplus, R_2= 2.05R_\oplus$, and $R_3= 2.54 R_\oplus$, to measure the sharpness of the features.  These bins are the positions of peaks and dip for group III stars (sun-like).\footnote{Our result is sensitive to the judicious choice of $R_2$, but less so on $R_1$ and $R_3$. If a bin other than that containing $R_2$ is adopted, the peak-to-dip ratio does not exhibit a peak at any value of $\beta$.}
% I can modify R1 and R3 majorly without changing result, but R2 must be exactly that bin
We measure the peak-to-dip ratio as $r_{1,\beta} = n(R=R_1)/n(R=R_2)$, $r_{2,\beta} = n(R=R_3)/n(R=R_2)$. The result of this procedure is presented in Fig. \ref{fig:gaia-shift-radius} and the value of $\beta$ that maximizes the peak to dip ratios lies near $\beta \approx 1$. The range of acceptable $\beta$ goes from $0.95$ to $1.4$, but in the remaining discussion, we adopt $\beta = 1$ for simplicity. There is no substantial change to the results when $\beta$ is varied in the above range.

% although the ratios get larger if we use the set (1.52,2.05,2.54), rather than (1.60,2.05,2.75)

To show that such a conclusion is not the result of random noise, we scramble the planet-star association and repeat the same procedure 100 times. None of these 100 runs return as large a peak-to-dip ratio as our real run for $\beta\approx1$.  In other words, the chance that random events can produce as sharp features as we do is
%(much) 
less than a percent.  

The scaled histograms, using $\beta=1$, are presented in Fig. \ref{fig:gaia-property-2}, alongside the original ones. This procedure not only lines up all features nicely among different groups of stars, it even reduces the radius dispersion within a given group. So the radius features caused by photoevaporation can be unified across different stars if we assume that the characteristic mass for super-Earths goes as $M_*^1$.

{ Prompted by one of the referees of this paper, I performed a new statistical analysis without breaking the planets into multiple groups. This procedure, detailed in the appendix, yields a similar result, $\beta \sim 0.92$.}

Alternative explanations to our above proposal are considered but are regarded as unlikely. First, we have addressed the issue of selection effect in the previous section. Second, planets orbiting around more massive stars tend to be hotter (at the same distance) and have larger atmospheric scale heights. This may shift the upper peak (slightly) upward, but not the lower peak (naked cores), nor the dip. Third, more massive stars tend to be born more recently and therefore are slightly more metal rich.  Could the observed progression be instead related to metallicity? We reject this hypothesis below.

\subsection{Core mass doesn't depend on metallicity}

Given that stellar metallicity should correspond to dust-to-gas ratio in the primordial proto-planetary disk, one may expect that planet masses correlate with stellar metallicity. Metallicity may also enter in the process of photoevaporation -- if planets around metal rich stars tend to have more metals in their gaseous envelopes, they may experience a different efficiency for photoevaporation.

Among FGK dwarfs that are investigated by the CKS group, stellar metallicities appear to rise with stellar mass. This is caused by the fact that more massive stars live shorter and therefore any observed ones are born more recently, after the Galaxy has become more enriched by metals. This correlation was noted by \citet{FultonPetigura}, and they suggest that metallicity may be (co-)responsible for their reported changes in the planet population, in addition to or instead of stellar mass.  Here, however, we argue that metallicity does not enter the picture.
%\subsection{``Heavy or Metal?''}
%Same question aksed by Gaidos+Mann14; Fulton+Petigura18

\begin{figure} \includegraphics[width=0.45\textwidth,trim=50 150 50 120,clip=]{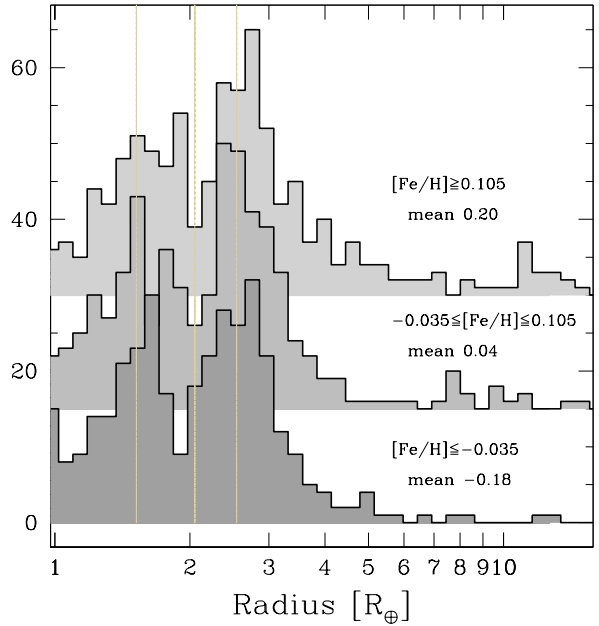} \caption{ Planet sizes have no correlation with host star metallicity.  There are $1141$ planets in GKS that overlap with the CKS sample and therefore have known host star metallicities. These are mostly around FGK dwarfs.  We separate the planets into three groups with increasing metallicity (ranges and mean values as marked) and display their respective histograms for the scaled radius ($R_p' =R_p (M_*/M_\oplus)^{-1/4}$). The average metallicity changes by a factor of $2.4$, yet there is little discernible progression in the distribution. There is, on the other hand, an excess of Jovian planets for the metal rich group { (as seen here at large radii)}.  }
\label{fig:gaia-property-3} 
\end{figure}

First, after the mass-dependence is scaled out ($\beta=1$ in eq. [\ref{eq:M0}]), one can show that the planet size distribution does not depend on metallicity (Fig. \ref{fig:gaia-property-3}). For FGK dwarfs that are in both the GKS and the CKS samples (1141 planets), there are no discernible trends as the average metallicity changes by a factor of $2.4$.

%Second, even without scaling out the mass dependency, one observes in Fig. \ref{fig:gaia-property-2} that %planets around groups I/II/III look different. The metallicity difference should be  small or negligible among %these three groups, as these stars outlive the Galaxy. 

% sm.s/gaia_metallicity
Second, among stars that reside in both the GKS and the CKS samples, the mean metallicity appears to rise only weakly with the stellar mass. For our division of 5 groups, the mean metallicities are, respectively, $<Fe/H> = -0.038$, $0.034, 0.036, 0.040, 0.023$. These changes are much smaller than the metallicity dispersion within a given group and are unlikely to be responsible for the systematic progression seen in Figs. \ref{fig:Rad-period-mass} and \ref{fig:gaia-property-2}.

It is now known that the occurrence rate of super-Earths is not strongly sensitive to stellar metallicity \citep{Buchhave,ZhuHuang}. This contrasts with the case of giant planets where the occurrence rate scales as $Z^2$ \citep{FischerValenti}, but is in line with our finding here that planet mass is independent of stellar metallicity. 

\section{Planet parameters -- Models of Photoevaporation}

Here, we extract information about the super-Earths by applying the model of photoevaporation. Based on the above discussion, we adopt a normal distribution for the planet mass, with a mean that scales with the stellar mass as $M_*^1$. And when comparing theoretical results against observations, we only compare the scaled planet radii, $R_p' = R_p (M_*/M_\odot)^{-1/4}$. This then allows us to leverage the entire GKS for our purpose, rather than having to restrict ourselves to one stellar group at a time.

\subsection{Characteristic Planet Masses}

\begin{figure} \includegraphics[width=0.49\textwidth,trim=20 170 50 90,clip=]{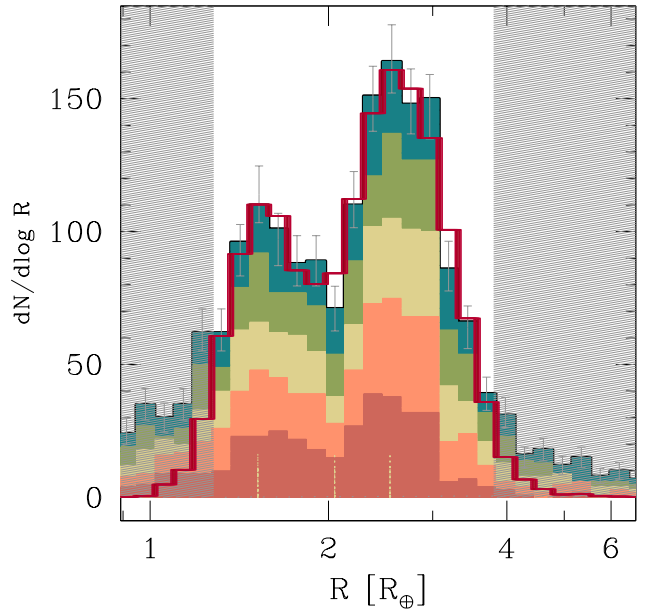} \caption{Comparing the observed 1D size distribution against our best-fit model. Here, all radii are scaled ($R_p'$). The shaded histogram is the observed one, with colors denoting contributions from groups I-V (color scheme as in Fig. \ref{fig:gaia-property-2}), and our best fit model (with $\chi_{\rm red}^2 = 0.9$) is shown as a heavy red histogram.  The un-shaded region is where we compare models against observations.
%There are $15$ radius bins that we fit for $6$ parameters, so the degree of freedom is $9$.  
}
  \label{fig:xvalley-shifted} \end{figure}

We simulate planets in the Gaia-Kepler sample ($1841$ planets). 
% I get 1841 = 2157-316, but merge_gaia.txt file reads to be 1976. don't know why a bit more
For each planet, we adopt its host star mass and its observed period as input. This saves us the labour of having to model the distributions of these two variables. We also do not need to consider transit probability.  We then produce 10 mock planets for each real planet, subject them to photoevaporation, measure their size distribution at a few Gyrs, and compare this against the observed 1-D distribution (Fig. \ref{fig:xvalley-shifted}). We have to introduce a number of parameters to describe this process, and we perform a $\chi^2$-minimization \citep[using the downhill simplex method,][]{Numerical} to find the best-fit parameters. The 2D (radius-period) distribution is not used in this procedure.

We assume the mock planets are drawn from a homogeneous population with core mass ($M_{\rm core}$), core radius ($R_{\rm core}$) and initial H/He mass fraction ($X$) as, \begin{eqnarray}
  {{dN}\over{d\log M_{\rm core}}}  & \propto & \exp\left[- { {(\log M_{\rm core} - \log M_c)^2}\over{2  \sigma_{\log M}^2}}\right]\, , \nonumber  \\
  R_{\rm core}  & = & 1.03 R_\oplus \left({{M_{\rm core}}\over{M_\oplus}}\right)^{1/4}\, , \nonumber \\
 {{dN}\over{d\log X}} & \propto & \exp\left[- {{(\log X - \log X_0)^2}\over{2 \sigma_X^2}}\right] \, ,  \nonumber \\
% I actually used a 3-power law now, better fit
%  {\rm Period distribution}\,\,
% {{dN}\over{d\log P}} & \propto & \begin{dcases} 
% {\rm constant} &  \text{for} \, P > 7.6 \left(M_*/M_\odot\right)^{1/4} \, \text{days} \\
%  P^{2.0} & \text{for}\, P \leq 7.6 \left(M_*/M_\odot\right)^{1/4} \, \text{days} \, .
%\end{dcases}
\end{eqnarray}
where $\sigma_{\log M}$ and $\sigma_X$ are the respective dispersions in logarithmic space.\footnote{We have experimented with Rayleigh distributions but see no dramatic changes.}  We explicitly set the characteristic mass $M_c = M_0 (M_*/M_\odot)$, where $M_0$ is the planet mass if the host star is solar. The planet core radius takes the form as in \citet{fortney2007a} to account for compression, and its normalization reflects our assumption that the planet bulk density, if it has a mass of $1 M_\oplus$, is $\rho_{1 M_\oplus} = 5.0 \g/\cm^3$ (the Earth has $\rho = 5.5 \g/\cm^3$), an assumption we justify later.

The above ansatz assumes that planet properties do not depend on orbital separation, an assumption we return to later.  It also assumes no correlation between core mass and envelope mass. A more nuanced treatment is desired but is not granted by the amount of data we have at hand.
% The third condition posits that all super-Earth cores are initially laced with H/He envelopes of a few percent %($3.2\%$ with a FWHM that runs from $2.5\%$ to $4\%$), again suggested by the observed size distribution %\citep{OwenWu13}. These values are determined by the (narrow) shape of the second peak.
% For instance, if the envelope mass fraction is higher, there would be too many large planets at large distances; %while if it is shifted to smaller values, there will be too many naked cores. And the spread is controlled by the %observed spread.  The period distribution is obtained by de-biasing the observed distribution, and a very weak %dependence is seen among the Gaia sample (hence the weak $(M_*/M_\odot)^{1/4}$ dependence on the %transition radius).
%The period distribution is aimed to emulate the Gaia sample, not necessarily the intrinsic one.
%also this is the one for planets with $R_p \in [1.28, 3.8 ] R_\oplus$.

With these initial conditions, we forward model the photoevaporation process as discussed in \citet{OwenWu17}. 
In particular, we adopt a high-energy luminosity $L_{\rm HE}$ that is $10^{-3.5} L_\odot (M_*/M_\odot)$ for the first $10^8$ yrs of the star's life, and let it decay as $t^{-1.5}$ afterwards. We include 
the difference in radius between the radiative photosphere and the transit radius, but do not include internal luminosities from the cooling core. And to allow for uncertainties in the photoevaporation model, 
we parametrize the energy efficiency of evaporation as 
\begin{equation}
\eta = \eta_0 \left({{v_{\rm esc}}\over{23 {\rm km}/{\rm s}}}\right)^{-\alpha_\eta}\, ,
\label{eq:eta}
\end{equation}
where $v_{\rm esc}$ is the surface escape velocity of the planet, and $23{\rm km/s}$ is the value for a super-Earth with $8 M_\oplus$ and $\rho_{1 M_\oplus} = 5\g/\cm^3$.  The value of $\eta$, together with that of $L_{\rm HE}$, determine the mass-loss rate. So we are not determining $\eta$ per se, rather the product of $\eta$ and $L_{\rm HE}$.  The case where $\alpha_\eta = 0$ corresponds to an 'energy-limited' scenario where a fixed fraction of the high-energy irradiation is employed to evaporate the envelope. This, as discussed in \citet{OwenWu17},is unlikely to be correct and physical arguments suggest that $\alpha_\eta > 0$.
% These values mainly manifest through the relatives heights of the two peaks and can so be chosen relatively %independently of other parameters. A pair of values, $\epsilon_0=0.2$ and $\alpha=1$, give reasonable fit in the 1-D size distribution (Fig. \ref{fig:xvalley-group3}).

After evolving a planet through thermal contraction and photo-evaporation, I account for detection completeness given its final size and orbital period, as well as its host star properties (\S \ref{subsec:selection}).
%.  to produce mock observations that account for both transit probability and recovery completeness for the %given star (with its associated radius and Kepler magnitude). 
As the completeness curves skirt  the main planet space, this does not much impact the results. I further include a Gaussian broadening of $7\%$ in planet radius to account for the error-spread in Gaia data. Binning the planets by their (scaled) radii as in Fig. \ref{fig:xvalley-shifted} allows us to calculate the $\chi^2$ difference between simulations and observations, where uncertainty arises from Poisson error for the latter. In detail, we focus only on planets with (scaled) radii between $1.28R_\oplus$ and $3.8 R_\oplus$. This contains $15$ bins in radius. With a total of $6$ parameters, this yields $9$ degrees of freedom, or the reduced $\chi_{\rm red}^2 = \chi^2/9$. Minimizing the difference leads us to a well defined global minimum at which $\chi_{\rm red}^2 = 0.9$. A single population of planets can explain the data, and their best-fit parameters are presented in Table \ref{tab:rho0}.  A few notable features. First, planet mass is found to be tightly clustered around $M_0 = 7.70 M_\oplus$ with a small dispersion ($\sigma_m= 0.3$, or a FWHM from $3$ to $14 M_\oplus$); the envelope mass also settles down to $X_0 = 2.6\%$ with a similarly small dispersion. The resulting 1-D histogram is presented in Fig. \ref{fig:xvalley-shifted}.

\begin{figure*} \includegraphics[width=0.95\textwidth,trim=0 0 0 0,clip=]{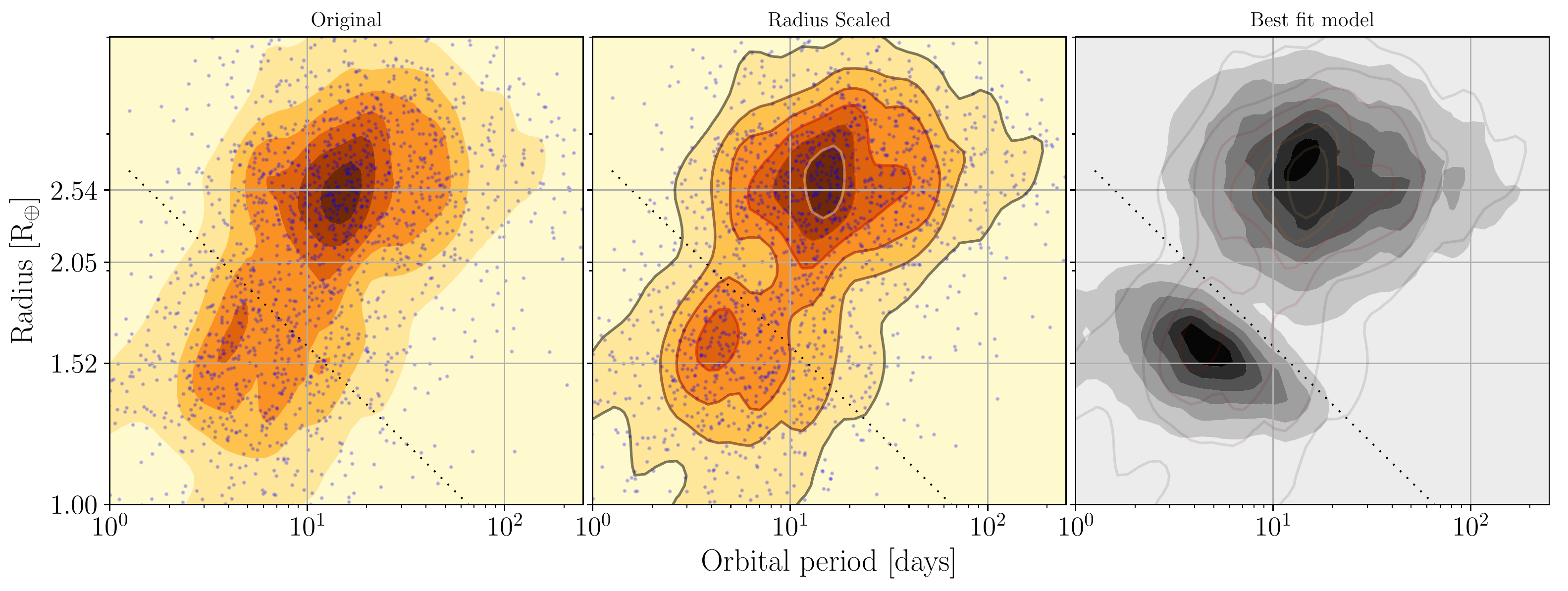} \caption{2D comparison between observation and best-fit model.  The left panel depicts planet radius vs. orbital period for the entire GKS. The middle panel is the same except all planet radii are substituted by the scaled values.  The right panel shows our best-fit simulations (also scaled radii), with grey contours reflecting that of the middle panel.  Horizontal lines indicate positions of peaks and valley in the 1D size distribution.  Notice that the evaporation valley is not clean, in either data or model.}  \label{fig:Rad-period-2} \end{figure*}

Our procedure only demands a best-fit in 1-D radius distribution. But as one can see in Fig. \ref{fig:Rad-period-2}, such a model also produces a satisfactory resemblance to the observed data in 2-D space.

% read in using sm.s/rho_dependence
\begin{deluxetable}{lccccccc}
\tablecaption{Best fit parameters for different values of the bulk density.
\label{tab:rho0}}
\tablehead{
$\rho_{1M_\oplus}$ & $M_0$ & $\sigma_{\log M}$ & $\eta_0$ & $\alpha_\eta$ & $X_0$ & $\sigma_{X0}$ & $\chi^2_{\rm red}$ \\
$[\g/\cm^3]$ & $[M_\oplus]$ & & & & & &  
}\\
\startdata
%       rho0        mass  sigma_mass         eta   eta_power          X0    sigma_X0
           2.0  &     3.08  &    0.54 &    0.008 &     0.52 &    0.003 &     0.20  & 1.05 \\
           3.0 &       4.79    &  0.43   &   0.05 &      0.42   &   0.012 &      0.23  & 0.90 \\
           4.0 &      6.10   &    0.33 &      0.11 &       0.40 &     0.022  &     0.35 & 0.91 \\
           $5.0^*$  &   7.70   &  0.29 &     0.17 &     0.42 &    0.026 &     0.37 & 0.90 \\
                         &  $\pm 1.5$ & $\pm$0.06 & $\pm$0.03 & $\pm$0.08 & $\pm$0.006 & $\pm$0.10 & \\
           6.0  &     9.43  &    0.28 &     0.29 &     0.54 &    0.030 &     0.33 & 0.75 \\
\hline \\
\multicolumn{8}{l} {$\gamma=1/4$} \\
      5.0& 8.73 & 0.32 & 0.16 & 0.42 & 0.016 & 0.41 & 0.72 \\
\multicolumn{8}{l}{$\gamma=1/2$}\\
      5.0 & 10.44 & 0.31 & 0.16 & 0.40 & 0.011 & 0.46 & 1.05 \\
\multicolumn{8}{l}{$\gamma=1$}\\
      5.0& 12.11 & 0.24 & 0.09 & 0.30 & 0.005  & 0.64 & 2.10\\
\enddata      
\tablecomments{
%For stars of different masses, values here refer to those normalized at $1 M_\odot$. The degree of freedom is $9$.
Model marked by an asterisk is our preferred model, for which we also present the $95\%$ confidence range for the estimated parameters.
% see lecture: http://astronomy.swin.edu.au/~cblake/StatsLecture3.pdf
% also, for 9 degrees of freedom, http://www.socscistatistics.com/pvalues/chidistribution.aspx
% gives that if chi^2 is 17, 5% likelihood,  this is equivalent to chi^2_{\rm red} increase by 1 from 1
The lower section lists models with a new parameter $\gamma$ (eq. [\ref{eq:adep}]).  }
%
%https://physics.stackexchange.com/questions/257772/least-squares-fitting-68-confidence-interval
% According to Bevington (Data Reduction and Error Analysis for the Physical Sciences), a single-parameter 68% % confidence interval is given by parameter values that increase the χ2 from min to min+1
\end{deluxetable}

\subsection{Core Composition?}

\begin{figure} \includegraphics[width=0.49\textwidth,trim=40 150 50 320,clip=]{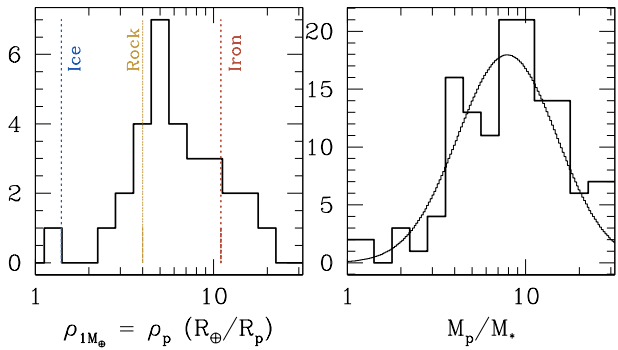} \caption{Left panel: scaled densities for transiting planets that also have RV measured masses. Here, we select only planets with observed radii between $1$ and $2 R_\oplus$, and orbital periods shortward of $30$ days. These are most likely naked cores. The scaled density is related to the real density as $\rho_{1 M_\oplus} = \rho (R_p/R_\oplus)^{-1}$. This is the density a planet of the same composition would have, if it had a mass of $1 M_\oplus$. The three coloured lines represent three pure compositions (ice, silicate and iron oxide). The observed ensemble peak near terrestrial composition (rock-iron mixture), or a scaled density of $\rho_{1M_\oplus} \sim 5\g/\cm^3$. Each density measurement comes with an error that ranges from $10\%$ to order unity. If we only select systems with more precise measurements, the sample size decreases but the same conclusion remains. Right panel: a histogram of RV masses for transiting planets that satisfy $R \leq 4 R_\oplus$. Our preferred model ($\rho_{1M_\oplus} = 5.0 \g/\cm^3$, see Table \ref{tab:rho0}) is shown as a black curve.  This agreement may be superficial as RV measurements are biased against detecting low-mass planets. }
\label{fig:planet-density} 
\end{figure}

\begin{figure*} \includegraphics[width=0.98\textwidth,trim=0 0 0 0,clip=]{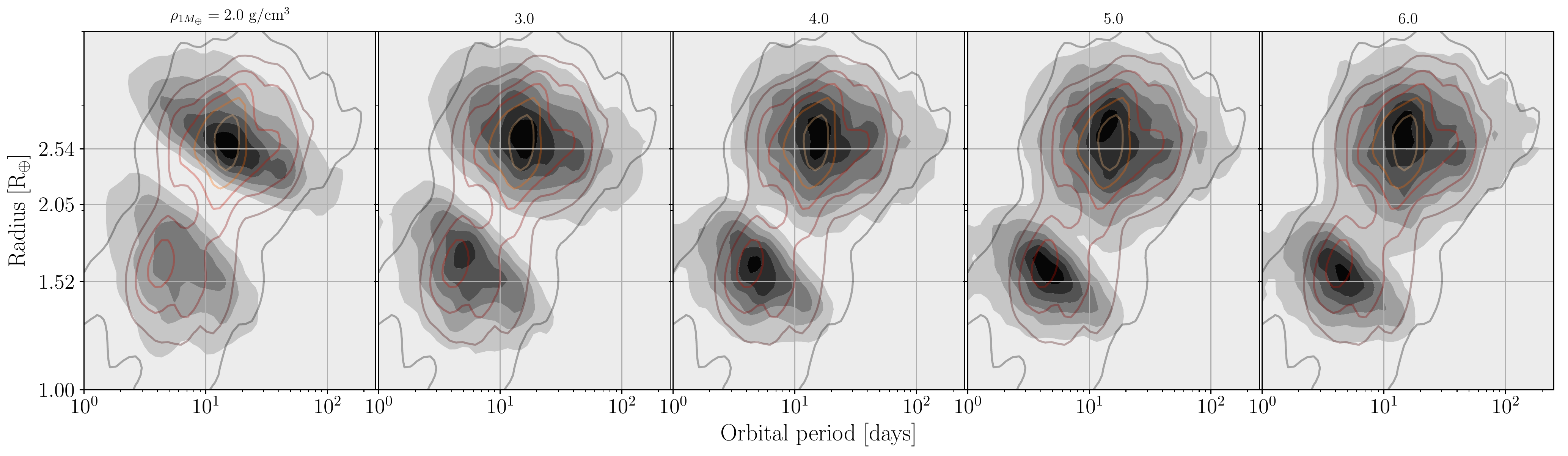} \caption{Bulk densities and 2D distribution. Best-fit models (for 1-D size distribution) are shown in the plane of (scaled) radius vs. orbital period, for various values of bulk density. Overlaid in { line}  contours are the actual observations (middle panel of Fig. \ref{fig:Rad-period-2}). Models with a lighter composition tend to leave a very clean 'evaporation valley' and tilt downward towards long periods. The fourth panel, {though not obviously the best fit here (the right three panels agree with the data equally well), is} our preferred model, and we give three arguments in text to justify our choice.}  \label{fig:Rad-period-rho0} \end{figure*}

In the above exercise, we fix $\rho_{1 M_\oplus}$, the bulk density for the same composition at $M = 1 M_\oplus$, to be $5.0 \g/\cm^3$. For context, a pure ice, pure rock, and pure iron composition should have $\rho_{1M_\oplus} = 1.4, 4.0$ and $11 \g/\cm^3$, respectively \citep{fortney2007a}, and the Earth, being a iron-rock mixture, has $\rho_{1 M_\oplus} = 5.5 \g/\cm^3$. So our choice implies a composition that is similar to that of the Earth.  We justify our choice here.

First, under a different assumption for the bulk density, we repeat the above exercise and obtain best-fit. These are listed in Table \ref{tab:rho0}, and $\chi_{\rm red}^2 \leq 1$ for all cases considered.  So the 1-D size distribution alone cannot distinguishes among the different compositions. How to break the degeneracy?

Among the listed solutions, one notices that, as the core bulk density decreases, so does the characteristic planet mass ($M_0$), in such a way that the lower peak at $R_1$ is preserved. In the mean time, the required photo-evaporation efficiency ($\eta_0$) drops dramatically, so as to preserve the large number of planets at the second peak despite the lower core mass.  The reverse happens when the bulk density increases. We can compare the best-fit evaporation efficiency against detailed models of photo-evaporation, where thermal calculations for the X-ray partial ionization region typically predict an evaporation efficiency of $\eta \sim 0.1$ \citep{OwenJackson, OwenWu13,Erkaev}. Such an efficiency disfavours models with too light or too dense a composition \citep{OwenWu17}, and favours our standard model ($5.0 \g/\cm^3$).
%, and corresponds roughly to the planet mass at size $R_p = 2 R_\oplus$ (the 'dip'). 
%In the meantime, mass dispersion ($\sigma_{\log M}$) rises, and envelope mass fraction ($X_0$) drops steeply.  In %particular, for our choice of bulk composition, $\eta \sim 0.18$ for mass

Other than theoretical support, we can also draw inspiration from the limited data at hand on planet density.  Fig. \ref{fig:planet-density} shows a collection of density measurements for RV planets that also transit. We restrict ourselves to planets that are likely naked-cores, that is, with (scaled) radii between $1$ and $2 R_\oplus$ and periods short-ward of $30$ days. We also scale the observed densities by the planet mass to obtain that of the same composition at $1 M_\oplus$. The resulting histogram suggests a peak at $\rho_{1 M_\oplus} \sim 5 \g/\cm^3$, more concentrated than when the (unscaled) physical densities are plotted.  Similarly, mass measurements of planets, naked cores or not, peak at $M \sim 8 M_\oplus$, the predicted location for terrestrial composition \citep[also see Fig. 10 of][]{mayor-small}. However, the mass distribution should be taken with caution as RV studies are known to be biased toward more massive planets.

Lastly, we can compare the 2D distribution of different best-fit models. Again, the one with terrestrial composition most resembles that of the observed one (Fig. \ref{fig:Rad-period-rho0}).

\subsection{Dependence on Orbital separation?}

We have adopted a model where planet properties (mass and envelope) are independent of the orbital separation. This gives adequate fit to the observations. However, it is reasonable to expect that planet formed (or parked) at different distances should have different core masses, and different models predict different distance scalings. How well can the current data constrain this?

We modify our model to include a radial dependence for the characteristic mass,
\begin{equation}
M_c = M_0  \left({{M_*}\over{M_\odot}}\right)\, \left({a \over{0.1 {\rm AU}}}\right)^\gamma\, .
\label{eq:adep}
\end{equation}
All other parameters, including the envelope mass, are assumed to be independent of the separation.  We then search for best-fit to the 1-D size distribution for a few values of $\gamma$. These are listed in Table \ref{tab:rho0}, together with their corresponding $\chi_{\rm red}^2$.  A rising $\gamma$ typically requires a larger $M_0$. The model with $\gamma=1$ fails to describe the 1D distribution ($\chi_{\rm red}^2 = 2.10$) but the ones with smaller $\gamma$ are successful. 

\begin{figure*} \includegraphics[width=0.95\textwidth,clip=]{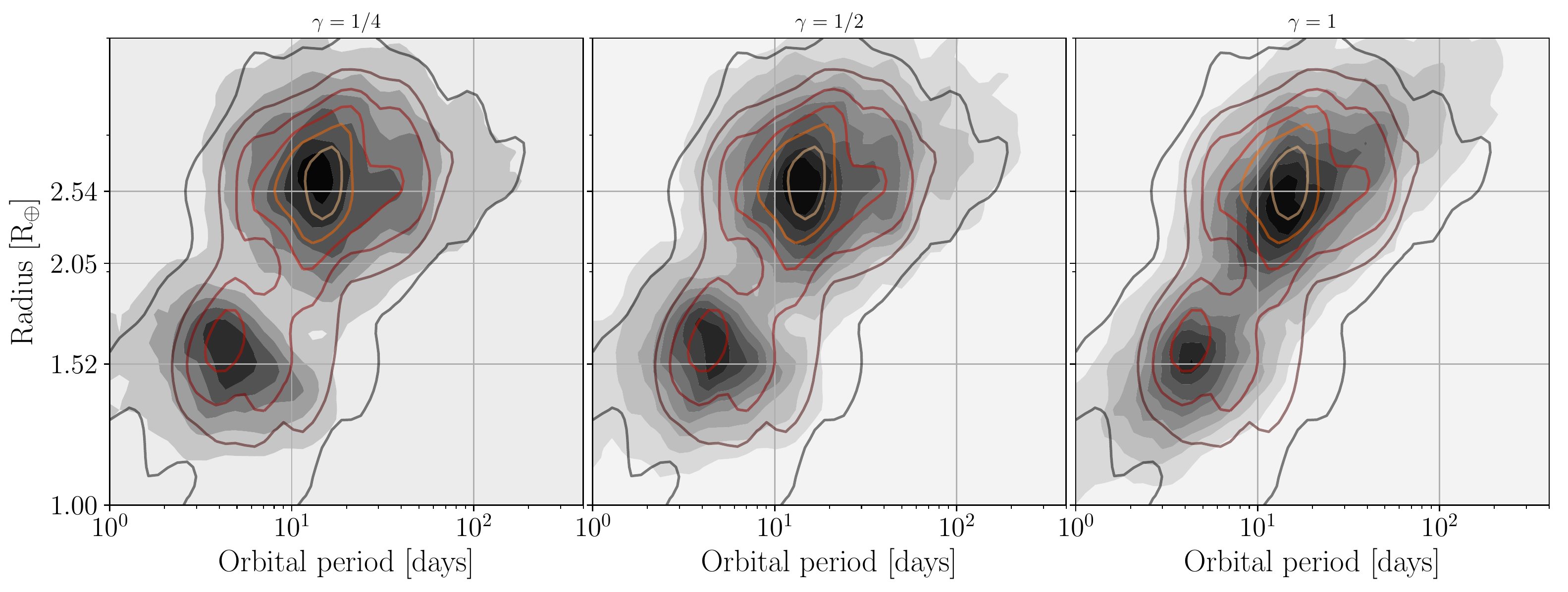} \caption{Constraining the radial dependence of core mass. Here, we compare different best-fit models (left: $\gamma = 1/4$; middle: $\gamma=1/2$; and right: $\gamma = 1$), against the observations (shown as { line} contours). The last model shows a upturn that is too steep to be compatible with current data, while other models produce acceptable resemblance to the observations, including the $\gamma=0$ model (4th panel in Fig. \ref{fig:Rad-period-rho0}). Visually, the $\gamma=1/2$ model produces the closest resemblance to data. }
  \label{fig:rad-period-beta} \end{figure*}

In Fig. \ref{fig:rad-period-beta}, we further compare these best-fit models in 2D space.  It is apparent that the $\gamma=1$ case is also ruled out by the 2D comparison, as it leads to too steep an upturn in planet size toward long periods. Models with smaller $\gamma$, on the other hand, appear compatible with the observations. Superficially, the $\gamma=1/2$ model looks most similar, but lacking a proper 2D statistical test, this is not quantifiable.

Aside from the technical difficulty of comparing models in 2D space, we are also hampered by the following issues. First, the {\it Kepler} sample contains too few long period planets ($\gg 10$ days) due to their smaller transit probabilities. So we are relying on a small batch of the total sample to discern any radial trend. Second, radii of planets at these long periods are a combined result of their core masses, their envelope masses, and their envelope scale heights. These may vary with separation in ways that conceal physical changes.
% RV planets median mass is: p<10 days, 7 M_E; p > 10 days, 11 M_E
The last issue can potentially be circumvented by radial velocity studies.
% -- there is some evidence from the RV sample that planet mass is rising with separation, but given the %well-known bias in RV detection, careful efforts have to be invested.

\section{Discussions}

The following expression summarizes our results succinctly. Masses of super-Earths peak at a mass ratio
\begin{equation}
\mu = {{M_p}\over{M_*}} \approx  2.5\times 10^{-5}   \, 
 \left({{M_*}\over{M_\odot}}\right)^a \, 
%\left({{\mu_d}\over{0.01}}\right)^b \, 
\left({{Z_*}\over{Z_\odot}}\right)^b \, 
\left({r\over{0.1 AU}}\right)^\gamma 
\, ,
\label{eq:constantmu}
\end{equation}
and that they are likely terrestrial in composition (based off Fig. \ref{fig:Rad-period-rho0}). Here, 
$a$ falls in a narrow range,  $a \in [-0.05,0.35]$ (Fig. \ref{fig:gaia-shift-radius}), while $ b \sim 0$ (Fig. \ref{fig:gaia-property-3}), and $\gamma < 1$ (we prefer $\gamma \sim 1/2$, Fig. \ref{fig:rad-period-beta}). %Here $\mu_d = M_d/M_*$ and $M_d$ is the mass of the proto-planetary disk at the time of formation. 
In the following, we discuss these results in more detail.

\subsection{Error in Owen \& Wu (2017)}

We adopt much of the same machinery as that in \citet{OwenWu17} to evolve the planet population. And the evaporation parameters in our standard model are similar to those used in that paper. However, our value for the core mass is a factor of $2$ higher than theirs, for sun-like stars. This difference stems not from the theory side, but from the data set we choose to compare against. In that paper, we took the 1D size distribution from Fig. 7 of \citet{Fulton}, after they have corrected for transit probability and pipeline completeness. As is also shown in \citet{FultonPetigura}, such a correction moves the two peaks downward to smaller radii. For instance, the small peak moved from $1.5$ to $1.3 R_\oplus$. This corresponds to a change in planet mass of $1.8$. Using this corrected size distribution led \citet{OwenWu17} to { infer a smaller core mass than is obtained here.}

This downward movement of the peaks is likely fictitious and is a result of over compensation. Small planets, especially those at long periods, due to their small detection probability, can strongly influence the corrected size distribution.  In this work, we perform a different procedure. We pass our model predictions through completeness correction to produce mock observations. In this way, we avoid the above issue.

\subsection{Other evidences of Mass dependence}

\begin{figure*} \includegraphics[width=0.95\textwidth,clip=]{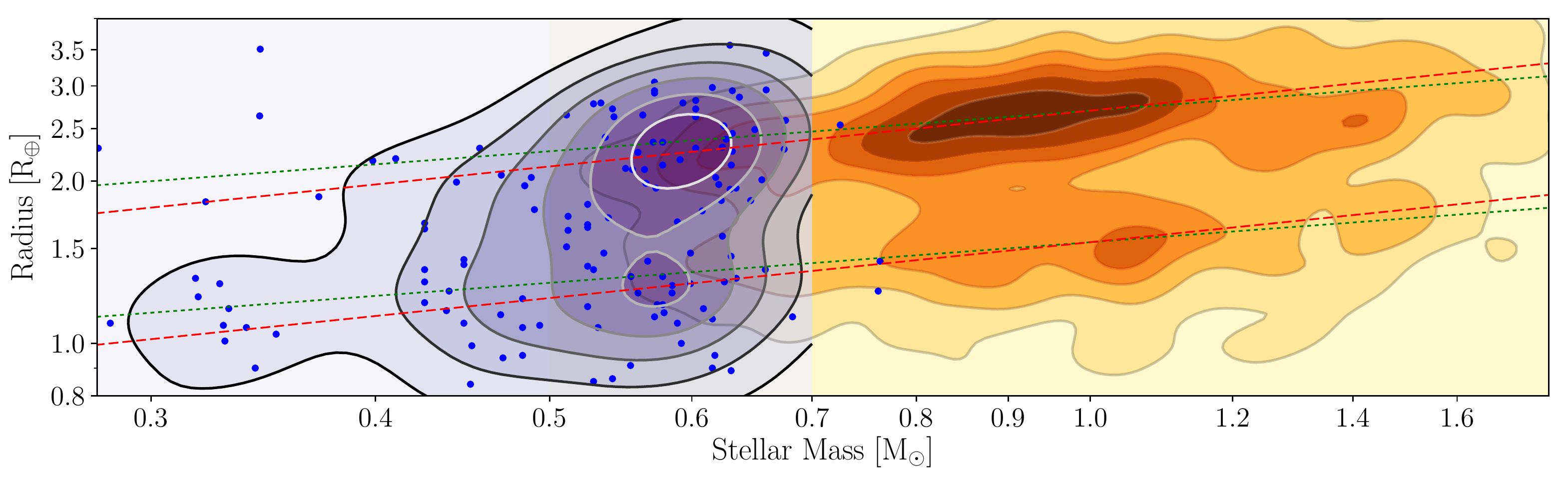} \caption{ Contours of planet occurrences in the stellar mass - planet radius plane. The orange one shows that of the GKS, an upward trend is clearly seen. This very trend appears to extend to late M-dwarfs as low in mass as $0.2 M_\odot$, as is shown by the blue contours and points \citep[data from][]{Newton,Dressing}.  Planets around late M-dwarfs also exhibit a radius gap that results from photoevaporation. The dotted green lines portray the $R_p \propto M_*^{1/4}$ scalings (which corresponds to $M_p \propto M_*$), while the dashed red lines that of $M_p \propto M_*^{11/8}$ (eq. \refnew{eq:bondim}). Both are compatible with data.}
\label{fig:rad-mass} \end{figure*}

The mass dependence that is apparent in the GKS extends to lower mass stars.  Two studies for planets around M-dwarfs from the {\it Kepler} and K2 mission have been published \citep{Newton,Dressing}. They carefully calibrated the M-dwarf spectra and obtained reliable radii for planet hosts with masses as low as $0.2 M_\odot$.  In Fig. \ref{fig:rad-mass}, I overplotted their results against the GKS (which only extends down to $0.5 M_\odot$). The clear progression seen in Gaia radii now extends down further. So for stars from $0.2 M_\odot$ to $2 M_\odot$, which includes the majority of stars in the universe, their super-Earths can be described by a single scaling law, $M_p \propto M_*^\beta$, with $\beta \approx 1$ (Fig. \ref{fig:rad-mass} shows two examples, $\beta=1$ and $\beta=11/8$).

Microlensing of planets around M-dwarfs provide mass measurements at larger separations (a few AUs).
Survey results have led practitioners in that field to propose that there is a 'break' in mass ratio around $\mu \sim 2\times 10^{-4}$ \citep{Suzuki,Udalski} . Adopting a radial dependence of $\gamma = 1/2$ and evaluate our eq. \refnew{eq:constantmu} at $r = 4$ AU, we obtain, intriguingly,
\begin{equation} 
\mu (r = 4 {\rm AU}, \gamma={1/2}) \sim 2\times 10^{-4}\, .  
\label{eq:microlensing} 
\end{equation}

To our knowledge, two previous papers have claimed to detect the dependence on stellar mass.  Using a set of planet radii determined by CKS spectroscopy and Gaia astrometry, as opposed to { only using} the Gaia data here, \citet{FultonPetigura} discovered that among the CKS FGK dwarfs, there is a progression of planet size with stellar mass. For their narrower mass range ($0.85$ to $1.2 M_\odot$), they found that the average planet radii increase by between $5\%$ to $13\%$.
% for the two radius ranges that they defined. 
If we substitute our scaling into their mass range, we expect a change of $9\%$, consistent with their claim.  They are uncertain as to the cause of the change, however, and suspect that the rising stellar metallicity may be partly responsible. In this work, we  exclude metallicity as a possible cause.

\citet{Pascucci} also noticed that the characteristic planet size rises with stellar mass, based on only KIC planet radii.\footnote{These have been shown to be much less accurate and may contain systematic errors\citep{Fulton}}. Using a mass-radius relation that is not motivated physically\footnote{And actually wrong -- planets at the upper peak and lower peak have identical mass, despite a factor of 2 difference in their sizes.}, they argued that planet mass function has a break at $\mu \sim 3\times 10^{-5}$. This is similar to the position of the 'peak' (not 'break') that we find here, relying on updated radii and a physical model. Our agreement, despite the very discrepant methodology, probably reflects the robustness of such a conclusion -- technical details matter, but the truth is already shining in the raw data.

\subsection{Why the universal scaling?}

So it appears that, despite the drastically different environments around stars from late M-dwarfs to early F-dwarfs, super-Earths are formed with a universal scaling (eq. \ref{eq:constantmu}). Why is this so?
The spread around the universal scaling is also tight.  Our best fit solution gives a Gaussian width of $\sigma_{\log M} = 0.3$ (in logarithm), or, the FWHM of the mass distribution runs from $\mu = 10^{-5}$ to $\mu = 5.3 \times 10^{-5} $ (corresponding to $3.5$ and $17.8 M_\oplus$ for a a sun-like host). This is a strikingly narrow range.\footnote{The narrowness can also be intuited from the 1D radius distribution. If the mass function has been a factor of 2 broader, the radius dip would have been wiped out.} Compare this to the following dynamic ranges: $2.3$ for the stellar mass in the GKS; a factor of $30$ in stellar luminosity; a factor of $\sim 10$ in stellar metallicity; a factor of $10-100$ in mass of proto-planetary disks...  

These likely suggest a single formation channel for these planets, in which stellar mass is the dominant variable (together with perhaps the orbital separation). Planet mass being independent of stellar metallicity (or, in association, the disk mass) likely indicates that the disk environments where these planets form do not have the same dust-to-gas ratio as that of a primordial disk. Lastly, there is a threshold below which planets will continue to grow, and above which growth is stalled. 
%\footnote{Anecdotally, similar mass ratios are observed for Jupiter's four large moons: Io, $\mu=4.7\times %10^{-5}$; Europa, $\mu=2.5 \times 10^{-5}$; Ganymede, $\mu=7.8\times 10^{-5}$; Callisto, $\mu=5.6\times %10^{-5}$).}
%Saturn's large moons: Titan: $\mu=2.4\times 10^{-4}$.
%Neptune's large moons: Triton $\mu=2\times 10^{-4}$.

In the following, I suggest that the simplest explanation for our results, is that super-Earths are at 'thermal mass'. Or, the masses of these planets are such that their Hill radii are equal to the scale height of their nascent gas disks.

% for sun, 10^{-8} M_\odot/yr equals a luminosity of 9e32 erg/s, compared this to solar luminosity of 4e33
% not much dimmer even in Pre main-sequence; 
% however, local deposition may mean the accretion is more important far-out?
% Dullemond: active disk gives T ~ r^{-3/4}, which is same as passive flat disk (irradiated by R_*/r)
%                    so also not a great fit for spectral index observed
Let us consider a passively irradiated (non-accreting) disk surrounding a star of mass $M_*$, radius $R_*$ and effective temperature $T_*$.  This leads to a minimum estimate for the disk scale height, and { we consider disks that are heated by active accretion further below.}
% may increase it moderately.  
Star rays hitting the slanted disk surface warms the interior temperature to \citep{ChiangGoldreich}
% I take alpha = d/dlnr (H/r) = 2/7 (H/r) = 2/7 * (H/H_g) (H_g/r) ~ 2/7 * 6 * (H_g/r)
%      I set H/H_g = 6 to be approximate, H/H_g = sqrt(2 ln (n_0/n_ph))
% knowing what the H/r dependence will be in the end
% also take H_g/r = c_s/v_kep, and c_s = sqrt(k_B T/mu/m_H), mu=2.3
% Chiang+Goldreich gives T=(alpha/4)^{1/4} (R_*/r)^{1/2} T_*, the extra factor of 2 account for green
% house effect; now alpha is a factor of 6 larger due to H/H_g, 
\begin{equation}
T (r) \approx 120 \K 
\left({r\over{1 AU}}\right)^{-3/7} \, 
\left( {{R_*}\over{R_\odot}}\right)^{4/7} \, 
\left({{M_*}\over{M_\odot}}\right)^{-1/7}\, 
\left({{T_*}\over{T_\odot}}\right)^{8/7}
\, .
\label{eq:Tflared}
\end{equation}
This is obtained by setting the optical photosphere to be some $6$ times larger than $H_g$ \citep{ChiangYoudin}.\footnote{The factor of $6$ applies to a MMSN-type disk. If the disk contains far more solid, this factor rises logarithmically and the gas scale height is moderately increased. } The above temperature is cooler than the temperature of an exposed blackbody.
% (which scales as $r^{-1/2}$). 
It also assumes that disk flaring dominates the slanted-ness in intercepting star rays. For the inner region, the finite angular size of the star dominates instead, in which case
%\footnote{ This expression has no dependence on stellar mass because it is mostly geometry 
% take alpha = 4/3pi (R_*/r), Chiang+Goldreich give T=(alpha/2)^{1/4} (R_*/r)^{1/2} T_*
\begin{equation}
%T (r) =   71 \K 
%\left({r\over{1 AU}}\right)^{-3/4} \, 
T(r) \approx 400 \K 
\left({r\over{0.1 AU}}\right)^{-3/4} \, 
\left( {{R_*}\over{R_\odot}}\right)^{3/4} \, 
\left({{T_*}\over{T_\odot}}\right)
\, .
\label{eq:Tanguar}
\end{equation}
Transition between the two regimes is approximately where the two scalings intersect, or,
$r_{\rm tran} \sim 0.2\, {\rm AU} (R_*/R_\odot)^{5/9}\, (M_*/M_\odot)^{4/9} \, (T_*/T_\odot)^{-4/9}$.
% Chiang & Goldreich take minimum of alpha, which occurs at ...
% or you can take the 3/4pi R_*/r = d/dlna (H/r)
Vertical gas scale height $H_g$ therefore satisfies
\begin{eqnarray}
& & {H_g\over r}= \nonumber \\
& &
\begin{dcases} 
& 0.013 \left({r\over{0.1 AU}}\right)^{1/8} \, 
\left( {{R_*}\over{R_\odot}}\right)^{3/8} \, 
\left({{M_*}\over{M_\odot}}\right)^{-1/2}\,
\left({{T_*}\over{T_\odot}}\right)^{1/2}\, ,\nonumber \\
& \hskip2.0in    r \leq r_{\rm tran} \\
&0.022 \left({r\over{1 AU}}\right)^{2/7} \, 
\left( {{R_*}\over{R_\odot}}\right)^{2/7} \, 
\left({{M_*}\over{M_\odot}}\right)^{-4/7}\,
\left({{T_*}\over{T_\odot}}\right)^{4/7}\, ,\nonumber \\
& \hskip2.0in   r \geq r_{\rm tran} \\
\end{dcases}
\label{eq:scaleheight}
\end{eqnarray}
To simplify the above scalings, we assume that pre-main-sequence stars follow the same mass-luminosity relation as that for ZAMS stars, $L_* \propto T_*^4 R_*^2 \propto M_*^4$. In this case, { $r_{\rm trans} \approx 0.2 {\rm AU} (R_*/R_\odot)^{7/9}$, or $0.3 {\rm AU}$ when $R_* = 1.7 R_*$, size of the Sun at 3 Myrs.}

Planet-disk interaction undergoes a transition at the so-called 'thermal mass', defined by the planet mass ratio at which its Hill radius exceeds the gas scale height. { For the above discussed passive disk, this is}
 \begin{eqnarray}
 && \mu_{\rm thermal}  = {{M_p}\over{M_*}} \equiv  3 \left({H_g\over r}\right)^3 
  \label{eq:bondim}\\
  & & \approx    \begin{dcases} 
& 6.3\times 10^{-6}   \left({r\over{0.1 AU}}\right)^{3/8} \, 
\left( {{R_*}\over{R_\odot}}\right)^{3/8} \,,
\hskip0.1in
   r \leq r_{\rm tran}, \nonumber \\
& 3.2 \times 10^{-5} \left({r\over{1 AU}}\right)^{6/7} \,, 
\hskip0.9in
r \geq r_{\rm tran}. \\
\end{dcases}
%\label{eq:bondim}
\end{eqnarray}
{ If stellar radii in the pre-main-sequence phase also scale as $M_*$, the inner branch of the thermal mass is $\mu_{\rm thermal} \propto M_*^{3/8}$, or $\gamma = 3/8$. This is actually slightly favoured by Fig. \ref{fig:rad-period-beta} over $\gamma =0$. It corresponds to $\beta = 1+3/8 =1.375$, and falls within our allowed range for $\beta$. }

{ Prompted by one of the referees, I also consider a model of the so-called active disk where viscous accretion dominates the heating of the inner disk. The fiducial disk in \citet{active} has an accretion rate of ${\dot M} = 10^{-8}M_\odot {\, \rm yr}^{-1}$, a turbulent viscosity parameter $\alpha = 10^{-2}$ and orbits around a star of mass $0.5 M_\odot$. Ignoring the stellar mass difference and taking a temperature profile from Fig. 1 of that paper, the inner region goes as
\begin{equation}
T \approx {\rm Max} \left[ 1600 \K, 235\K \left({r\over{1 {\rm AU}}}\right)^{-0.93}\right]\, .
\end{equation}
Here, $1600\K$ corresponds to the dust sublimation temperature and the cap at this temperature reflects the changes in gas opacity when sublimation occurs.  Outside a couple AU, active heating is sub-dominant and the passive disk model applies again. This model of an actively accreting disk represents a likely upper limit to the disk scale height, an opposite limit to the passive disk.}\footnote{However, unlike the passive disk, which can explain the observed stellar mass dependence, the active disk does not.}

\begin{figure*} 
\includegraphics[width=0.49\textwidth,clip=]{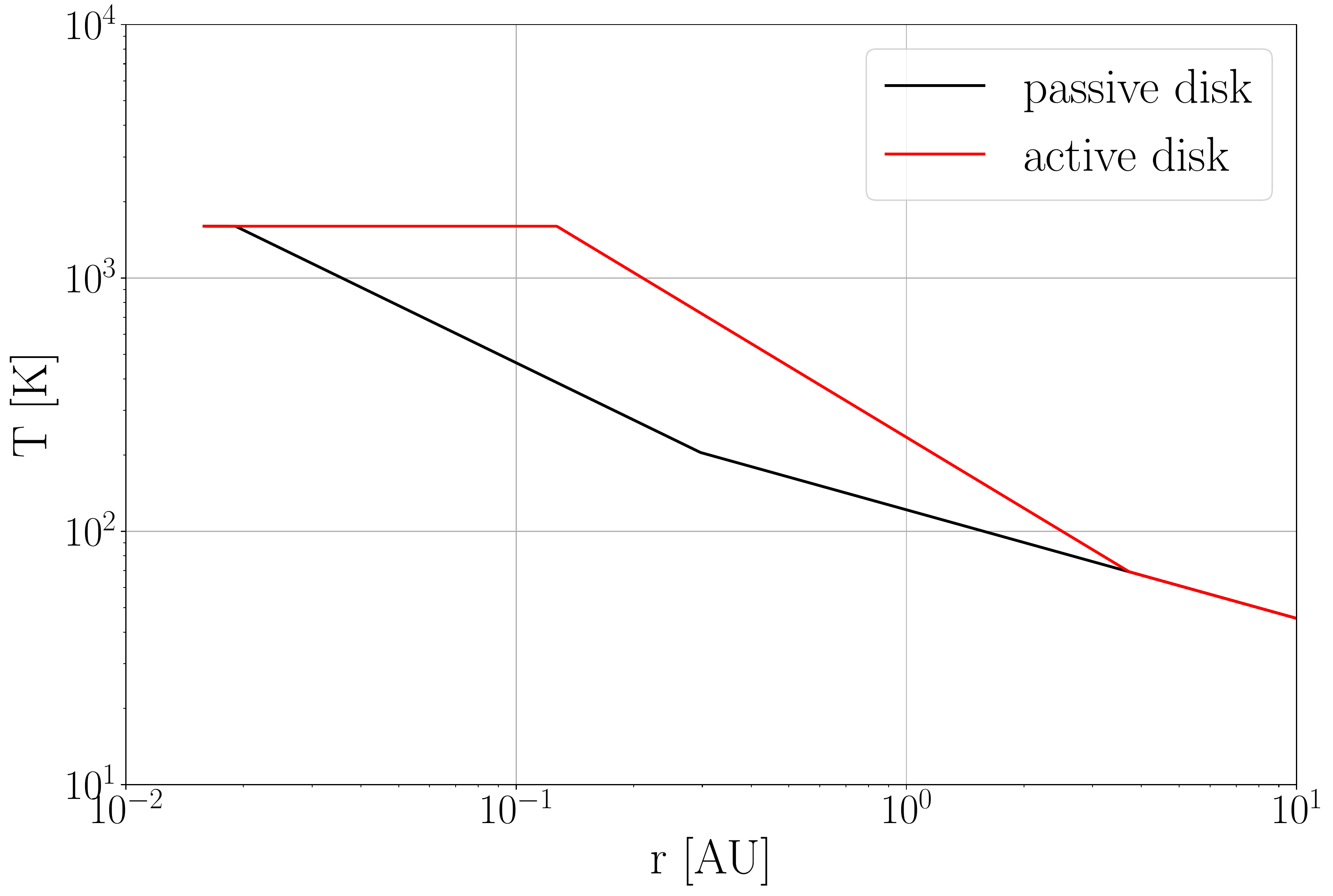} 
\includegraphics[width=0.49\textwidth,clip=]{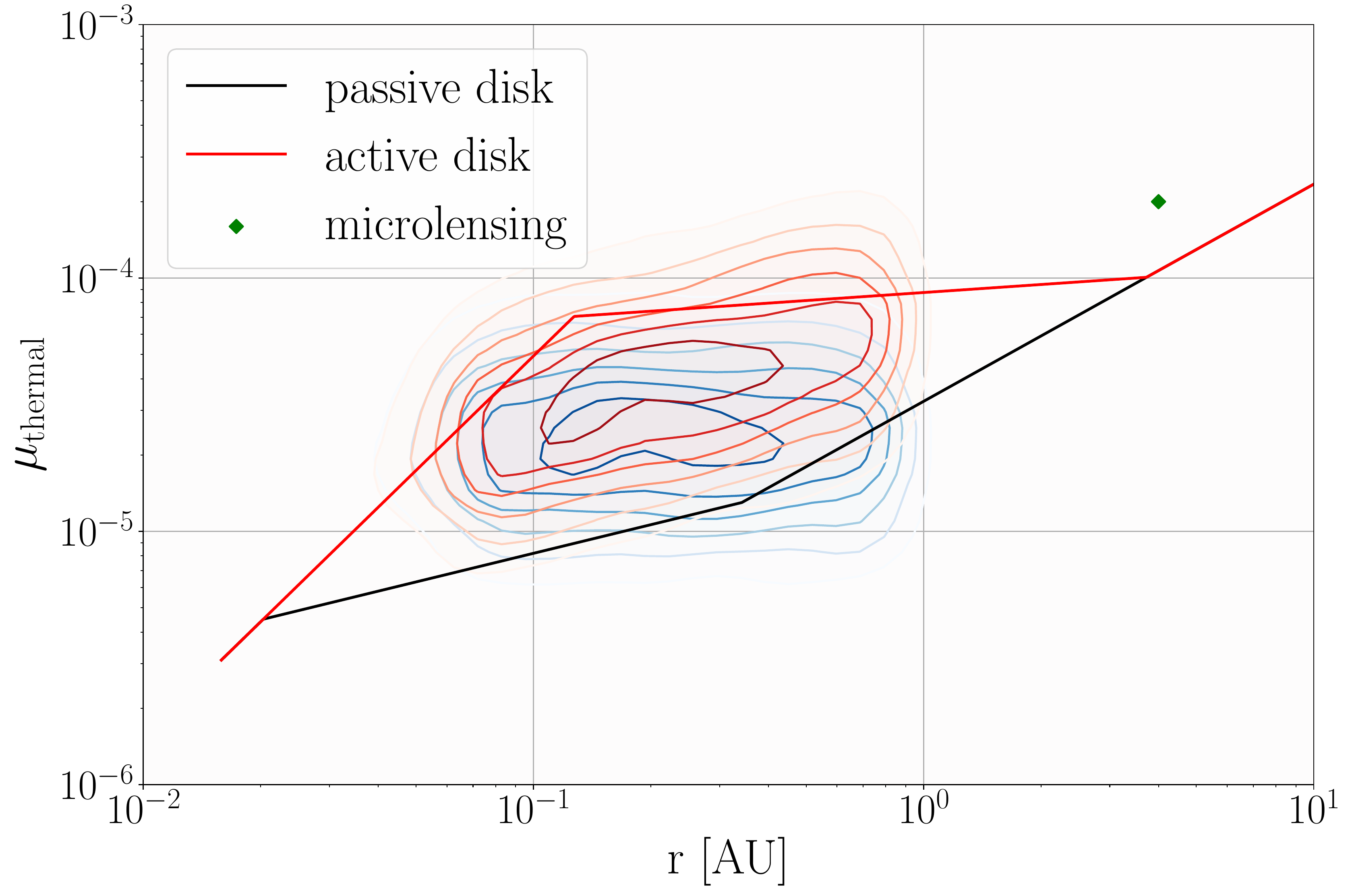} 
\caption{{ The left panel depicts the temperature profile for a passive disk (black line, with $R_* = 1.7 R_\odot$, and $T_* = 4500\K$) and for an actively accreting disk % the radius is taken from Fig. 2 of Gough1980
\citep[red line, a T-Tauri model from][]{active}, as functions of orbital separations. These likely represent the upper and lower limits for disk temperatures. 
Thermal masses corresponding to these two models are plotted as lines on the right panel, while the
red and blue contours display our best-fit solutions for super-Earth masses when $\gamma = 3/8$ and $0$, respectively (mean and dispersion taken from Table \ref{tab:rho0}).
 %There is a bent from the inner to the outer region, corresponding to the transition in disk flaring. The 
The observed mass distribution appears to lie in-between the thermal masses
expected for the two disk models.}
The green diamond is the 'mass-break' reported by microlensing surveys.  }
\label{fig:plot-planet} 
\end{figure*}

Fig. \ref{fig:plot-planet} illustrates the thermal mass for both the inner and the outer regions, around a solar-type star{, considering either an active or a passive disk}. We also overplot our best-fit models (represented by two, one with $\gamma=0$, the other $\gamma=3/8$).  Both solutions 
%fall close to (and slightly above) the theoretical thermal mass. 
{ fall within the values spanned by an active disk and a passive disk. Or, the observed planet mass distribution appear to be consistent, in both scaling and magnitude, with the expected thermal masses for proto-planetary disks.}

What is the significance of such an interpretation?

For a planet more massive than the thermal mass, \citet{LinPapaloizou} argued that it excites strong tidal shocks which can remove material from its vicinity to open a gap.  Disk viscosity counter-acts to smooth out the gap but so long as the viscosity parameter falls below  \citep[e.g., eq. (3) in][]{Crida}
% assume $\Delta_m = R_Hill$
\begin{equation}
\alpha_{\rm crit} \approx 0.04 \left({\mu\over{2.5\times 10^{-5}}}\right) \left({{H_g/r}\over{0.013}}\right)^{-2} \, ,
\label{eq:alpha}
\end{equation}
the planet wins and a gap is opened. Numerical simulations show that $\alpha$ in active MRI regions typically have values of $\alpha \sim 10^{-2}$, and it falls below that by orders of magnitude in disk mid-plane if MRI is not operative. Our super-Earths should be able to open gaps.

Planets that open gaps may migrate at a speed reduced from the rapid Type I migration rate\citep{LinPapaloizou}. The reduction depends on the depth and width of the gap \citep{Duffell14}.
%this paper fig. 3 shows that the migration rate in Type II is a function of Mach number, though higher 
% mach number (smaller H/R) also means a lower viscous rate for the same alpha
In addition, gap opening planets can potentially regulate their own growth, either by creating a pressure maximum to stall the inward spiral of 'pebbles' \citep{pebble,InsideOut}, or by interacting with larger planetesimals embedded in gaseous disks.  There is a large body of studies on gap-opening planets, but most have so far been limited to Jovian mass planets in disks with large aspect ratios, and much of their conclusions (e.g., regarding gas flow across gap, migration rate, eccentricity excitation,...) may have to be modified for the case of low-mass planets in thin disks.

There is another notable feature in our best-fit model.  The inferred envelope mass fraction  peaks tightly at $2\%$, with a FWHM that stretches from $\sim 1\%$ to $4\%$.\footnote{Again, this can be directly observed in the 1D size distribution -- notice the sharp drop-off of planets beyond $\sim 3 R_\oplus$.}  For comparison, the expected mass for an adiabatic envelope surrounding a $8M_\oplus$ planet still embedded in the disk should be $10-50\%$ \citep[e.g.][]{rafikovgas}.\footnote{But the assumption of 'embedded' is probably broken. See above.}  \citet{OwenWu16} discussed a mechanism (which they termed 'boil-off') in the early stage of a planet's life that quickly removes most of this envelope and pares it down to a few percent. But it is unclear whether this completely explains the observed narrow concentration of envelope masses.

\section{Conclusion}

Photo-evaporation gives us a sensitive tool, in addition to RV and TTV methods, to measure planet mass as a population. We find that a narrow mass function which rises roughly linearly with the stellar mass best explains the observed size distribution of {\it Kepler} planets. Our results are captured by the expression in eq. \ref{eq:constantmu} and here is a brief re-cap: \begin{itemize}

\item planet mass rises with stellar mass as $M_*^\beta$, with $\beta \in [0.95,1.40]$;

\item a characteristic planet-to-star mass ratio $\mu \sim 2.5\times 10^{-5}$ (i.e., planet mass $8 M_\oplus$ around a Sun-like star);

\item a  narrow spread of $\sim 0.3$ in logarithmic mass;

\item no evidence of metallicity dependence;

\item composition likely terrestrial-like;

\item planet mass possibly exhibits a radial dependence of  $r^\gamma$ with $\gamma \in [0,1/2]$;

\item all planets initially covered with H/He envelopes  $\sim 2\%$ in mass fraction;

\item planet masses lie near 
%or just above 
the local thermal mass.

\end{itemize}
These conclusions should be refined or further tested. First, in this work, we have used Gaia stellar radius as a proxy for stellar mass. This is strictly correct only for stars close to the ZAMS.\footnote{More massive stars evolve faster. This may explain why in the right panel of Fig. \ref{fig:gaia-property-2}, even after mass scaling, sizes of planets around more massive stars still appear more broadly distributed.}  More accurate stellar parameters will not only update the results here, but may also discern more hidden patterns. Second, as shown in Fig. \ref{fig:rad-mass}, planets around low-mass M-dwarfs are unique lever-arms in ascertaining the planet-to-star correlation. TESS and other surveys may provide a large sample of such objects.  Third, it is important to constrain mass dependence on orbital separation, as it is pivotal in identifying the correct formation channel. One venue is to expand the planet sample by establishing the planet nature of many of the currently so-called candidates; the other is to have precise radial velocity follow-up of low-mass planets across a range of separations.
%They can be included in the analysis, once their stellar parameters are firmly established.
% for the moment, my gaia radius for them are wierd...
Fourth, a more robust model for photoevaporation, capturing the dependency of the evaporation efficiency on various planet parameters, would be instrumental in determining the planet composition. Lastly, how a low-mass gap-opening planet interacts with the gas and the solid disks should be investigated.  

A few parting thoughts. 

Within the $400$ day period range probed by the {\it Kepler} mission, many systems show multiple transiting planets \citep{Lissauer,Fabrycky}. The average multiplicity has been inferred to be $\sim 3$ \citep{Zhu18}.  Taken together with the high mass of the super-Earths, it is tempting to think that the material for these super-Earths come drifting in from larger distances. However, such a scenario would have to accommodate the fact that many of the systems harbouring super-Earths ($\sim 30\%$ in average) also host cold Jupiters at large distances \citep{ZhuWu18}. 
%To firmly determine the composition of super-Earths may shed some light on this issue.

Both super-Earths and Jupiters, the two big classes of planets that we now know, are capable of gap-opening. This raises the suspicion that all planets formed in gaseous disks have to be gap-opening to survive.{ In other words, planets in the outer region, where the thermal mass is higher (Fig. \ref{fig:plot-planet}), ought to be more massive than the super-Earths we discuss here and are closer in mass to Saturn.  This perhaps explains the 'mass break' at $\mu \sim 2\times 10^{-4}$ from microlensing surveys.}

%\footnote{The solar system ice giants may be stalled in resonance by Jupiter/Saturn.}

\acknowledgements 

I wish to thank my two referees, as well as Yoram Lithwick, James Owen for discussions and technical help, and Wei Zhu for providing helpful feedback on the draft. This work makes use of the NASA Exoplanet archive, and results from the NASA Kepler mission, the ESA Gaia mission and the California-Kepler survey.  I acknowledge NSERC for a research grant.

%\bibliographystyle{apj}
%\bibliography{ref}

\begin{appendix}

\section{Stellar Dependence From a Gaussian Mixture Analysis}
{
Prompted by one of the referees, here I try to establish the dependence of planet radius on stellar mass using a method commonly used in statistical studies, that of a multi-component Gaussian mixture model. With such an analysis, we can avoid breaking the sample up into  somewhat arbitrary groups, as is done in the main text.

Using Mclust \citep{mclust}, a contributed package to the R language, we model the clustering of data in the 3-D space spanned by planet radius, planet period and stellar radius (all in logarithm). We insist that there are three or fewer Gaussian components in the data, and find that the best model (in terms of Bayesian Information Criterion) consists of 2 components, one clearly corresponding to the super-Earths, and the other sub-Neptunes. The mean planet radii, orbital periods and stellar radii for the two components are, respectively, $(1.5 R_\oplus, 5.68 {\rm days}, 0.99 R_\odot )$, $(2.63 R_\oplus, 18.71 {\rm days},0.98 R_\odot)$. The two populations share the same stellar properties, but differ in size and period, as is expected. These are shown in Fig. \ref{fig:R-experiment}. The ellipses that describe the Gaussian models are clearly inclined for both populations, indicating a dependence of planet radius on stellar radius (a proxy for stellar mass).

\begin{figure} 
\includegraphics[width=0.45\textwidth,trim=0 0 20 20 20,clip=]{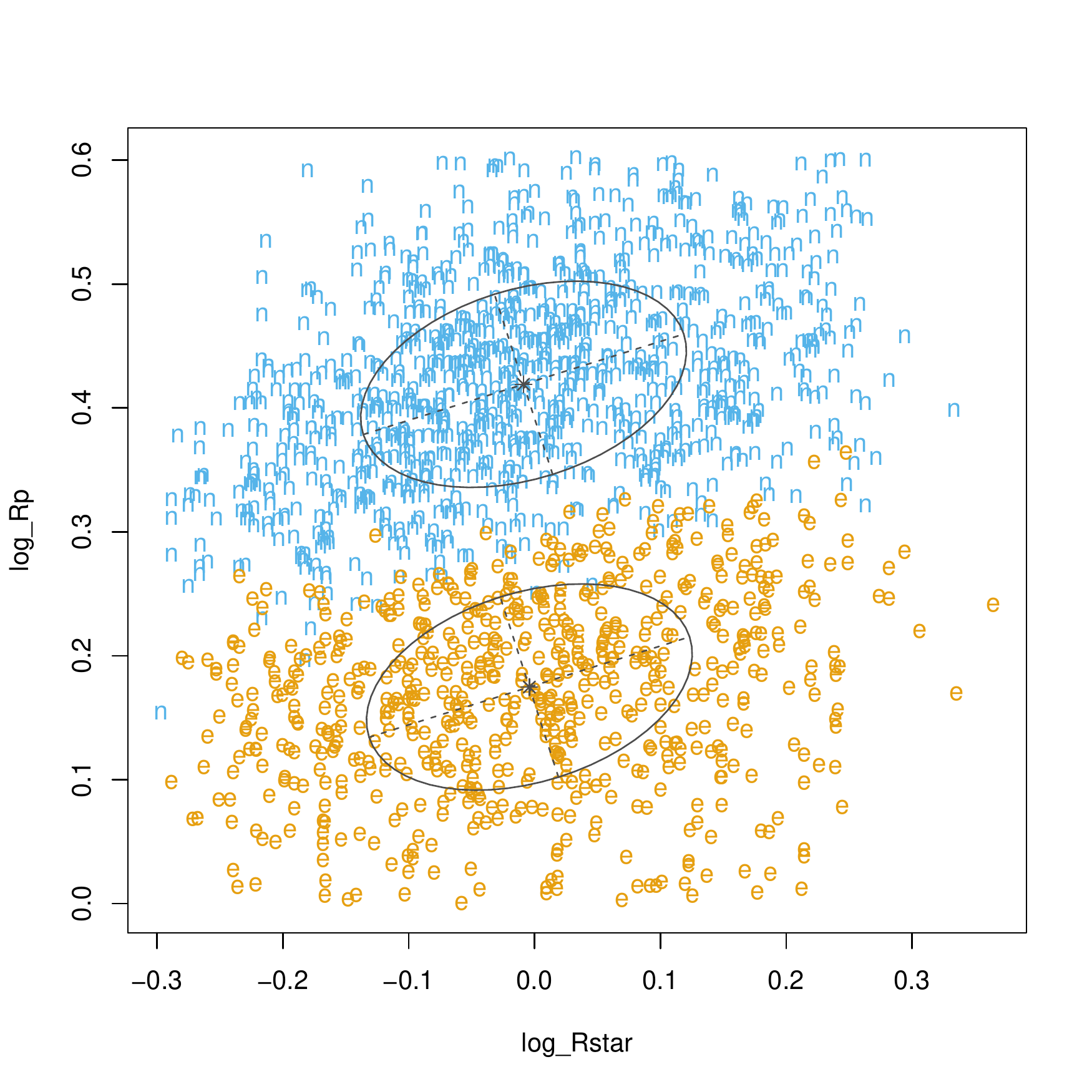} \caption{Cluster analysis using the Mclust package in R programme. The Gaussian cluster marked as brown-yellow points (labelled as 'e') corresponds to super-Earths, while the blue points labelled as 'n' to sub-Neptunes. The axis are logarithmic in stellar radius (horizontal) and planet radius (vertical). Mean and variance of the Gaussian components are indicated by the two ellipses.
}
\label{fig:R-experiment}
\end{figure}

% read ../README_statistics for detail
The covariance matrix for both components give the same picture, with the correlation between planet radius and stellar radius having a value of $0.33$, much stronger than correlations between the other pairs. 
% between radius-period, it is 0.08, between period and stellar radius it is $0.045$.
Moreover, decomposing the variations into basis vectors, I find that the principle component along which most of the data variations are explained is along the direction of $R_p \propto R_*^{0.23}$, or $\beta \approx 0.92$.  This confirms results from the more primitive analysis in the main text, where we find $\beta \approx 1$ (Fig. \ref{fig:gaia-shift-radius}).}

\end{appendix}

\end{document}